\documentclass[acmlarge]{acmart}
\AtBeginDocument{%
  }

\setcopyright{cc}
\setcctype{by}
\acmJournal{IMWUT}
\acmYear{2026} \acmVolume{10} \acmNumber{3} \acmArticle{98}
\acmMonth{9} \acmDOI{10.1145/3831629}

\newcommand{\add}[1]{\textcolor{black}{#1}}
\newcommand{\delete}[1]{}
\newcommand{\modify}[1]{\textcolor{black}{#1}}

\usepackage{hyperref}
\usepackage{graphicx}
\usepackage{booktabs}
\usepackage{caption}
\usepackage{subcaption}
\usepackage{enumitem}
\usepackage{arydshln}

\usepackage{multirow}
\usepackage{multicol}
\usepackage{color}
\usepackage{colortbl}
\definecolor{Gray}{gray}{0.9}
\newcolumntype{g}{>{\columncolor{Gray}}c}
\usepackage{graphicx}
\usepackage{makecell}

\usepackage{subcaption}

\usepackage{array}
\newcolumntype{C}[1]{>{\centering\arraybackslash}p{#1}}

\usepackage{etoolbox}

\AtBeginEnvironment{thebibliography}{%
  \interlinepenalty=10000
  \clubpenalty=10000
  \widowpenalty=10000
}

\usepackage{xcolor}
\usepackage{hyperref}

\newcommand{\blueurl}[1]{\textcolor{blue}{\url{#1}}}

\newcommand{\orcidname}[2]{%
  \href{https://orcid.org/#1}{#2}}

\makeatletter
\renewcommand{\@fnsymbol}[1]{%
  \ifcase#1\or
    \TextOrMath{\textdagger}{\dagger}\or
    \TextOrMath{\textasteriskcentered}{*}\or
    \TextOrMath{\textdaggerdbl}{\ddagger}\or
    \TextOrMath{\textsection}{\mathsection}\or
    \TextOrMath{\textparagraph}{\mathparagraph}\or
    \TextOrMath{\textbardbl}{\|}\or
    \TextOrMath
      {\textasteriskcentered\textasteriskcentered}{**}\or
    \TextOrMath
      {\textdagger\textdagger}{\dagger\dagger}\or
    \TextOrMath
      {\textdaggerdbl\textdaggerdbl}{\ddagger\ddagger}\else
    \@ctrerr
  \fi
}
\makeatother

\begin{document}

\title[Earinter]{Earinter: A Closed-Loop System for Eating Pace Regulation with Just-in-Time Intervention Using Commodity Earbuds}


\author{Jun Fang}
\authornote{Co-author equal contribution.}
\orcid{0009-0001-2614-8674}
\affiliation{
  \institution{Tsinghua University}
  \city{Beijing}
  \state{Beijing}
  \country{China}
}
\email{fangy23@mails.tsinghua.edu.cn}

\author{Ka I Chan}
\authornotemark[1]
\orcid{0009-0001-4560-1702}
\affiliation{
  \institution{University of Michigan}
  \city{Ann Arbor}
  \state{Michigan}
  \country{United States}
}
\email{chankai@umich.edu}

\author{Xiyuxing Zhang}
\authornotemark[1]
\orcid{0009-0002-9337-2278}
\affiliation{
  \institution{Tsinghua University}
  \city{Beijing}
  \state{Beijing}
  \country{China}
}
\email{zxyx22@mails.tsinghua.edu.cn}

\author{Yuntao Wang}
\authornote{Corresponding author.}
\orcid{0000-0002-4249-8893}
\affiliation{
  \institution{Tsinghua University}
  \city{Beijing}
  \state{Beijing}
  \country{China}
}
\email{yuntaowang@tsinghua.edu.cn}

\author{Mingze Gao}
\orcid{0009-0001-0792-7605}
\affiliation{
  \institution{Tsinghua University}
  \city{Beijing}
  \state{Beijing}
  \country{China}
}
\email{gaomingze2022@gmail.com}

\author{Leyi Peng}
\orcid{0009-0003-6827-6820}
\affiliation{
  \institution{Yale University}
  \city{New Haven}
  \state{Connecticut}
  \country{United States}
}
\email{leyi.peng@yale.edu}

\author{Jiajin Li}
\authornote{Visiting student at the time of this work.}
\orcid{0009-0001-5723-5383}
\affiliation{
  \institution{Tsinghua University}
  \city{Beijing}
  \state{Beijing}
  \country{China}
}
\email{lijiajin0516@gmail.com}

\author{Zihang Zhan}
\orcid{0009-0005-1001-0573}
\affiliation{
  \institution{Tsinghua University}
  \city{Beijing}
  \state{Beijing}
  \country{China}
}
\email{zhanzh22@mails.tsinghua.edu.cn}

\author{Zhixin Zhao}
\orcid{0009-0007-1726-2081}
\affiliation{
  \institution{Tsinghua University}
  \city{Beijing}
  \state{Beijing}
  \country{China}
}
\email{zhaozhix22@mails.tsinghua.edu.cn}

\author{Yuanchun Shi}
\orcid{0000-0003-2273-6927}
\affiliation{
  \institution{Tsinghua University}
  \city{Beijing}
  \state{Beijing}
  \country{China}
}
\email{shiyc@tsinghua.edu.cn}

\authorsaddresses{%
Authors' Contact Information:
\orcidname{0009-0001-2614-8674}{Jun Fang},
Tsinghua University, Beijing, Beijing, China, fangy23@mails.tsinghua.edu.cn;
\orcidname{0009-0001-4560-1702}{Ka I Chan},
University of Michigan, Ann Arbor, Michigan, United States, chankai@umich.edu;
\orcidname{0009-0002-9337-2278}{Xiyuxing Zhang},
Tsinghua University, Beijing, Beijing, China, zxyx22@mails.tsinghua.edu.cn;
\orcidname{0000-0002-4249-8893}{Yuntao Wang}
\space (corresponding author),
Tsinghua University, Beijing, Beijing, China,
Qinghai University, Xining, Qinghai, China, yuntaowang@tsinghua.edu.cn;
\orcidname{0009-0001-0792-7605}{Mingze Gao},
Tsinghua University, Beijing, Beijing, China, gaomingze2022@gmail.com;
\orcidname{0009-0003-6827-6820}{Leyi Peng},
Yale University, New Haven, Connecticut, United States, leyi.peng@yale.edu;
\orcidname{0009-0001-5723-5383}{Jiajin Li},
Tsinghua University, Beijing, Beijing, China, lijiajin0516@gmail.com;
\orcidname{0009-0005-1001-0573}{Zihang Zhan},
Tsinghua University, Beijing, Beijing, China, zhanzh22@mails.tsinghua.edu.cn;
\orcidname{0009-0007-1726-2081}{Zhixin Zhao},
Tsinghua University, Beijing, Beijing, China, zhaozhix22@mails.tsinghua.edu.cn;
\orcidname{0000-0003-2273-6927}{Yuanchun Shi},
Tsinghua University, Beijing, Beijing, China,
Qinghai University, Xining, Qinghai, China, shiyc@tsinghua.edu.cn.%
}

\renewcommand{\shortauthors}{Fang et al.}

\begin{abstract}

\begin{figure}[h]
    \centering
    \includegraphics[width=.95\linewidth]{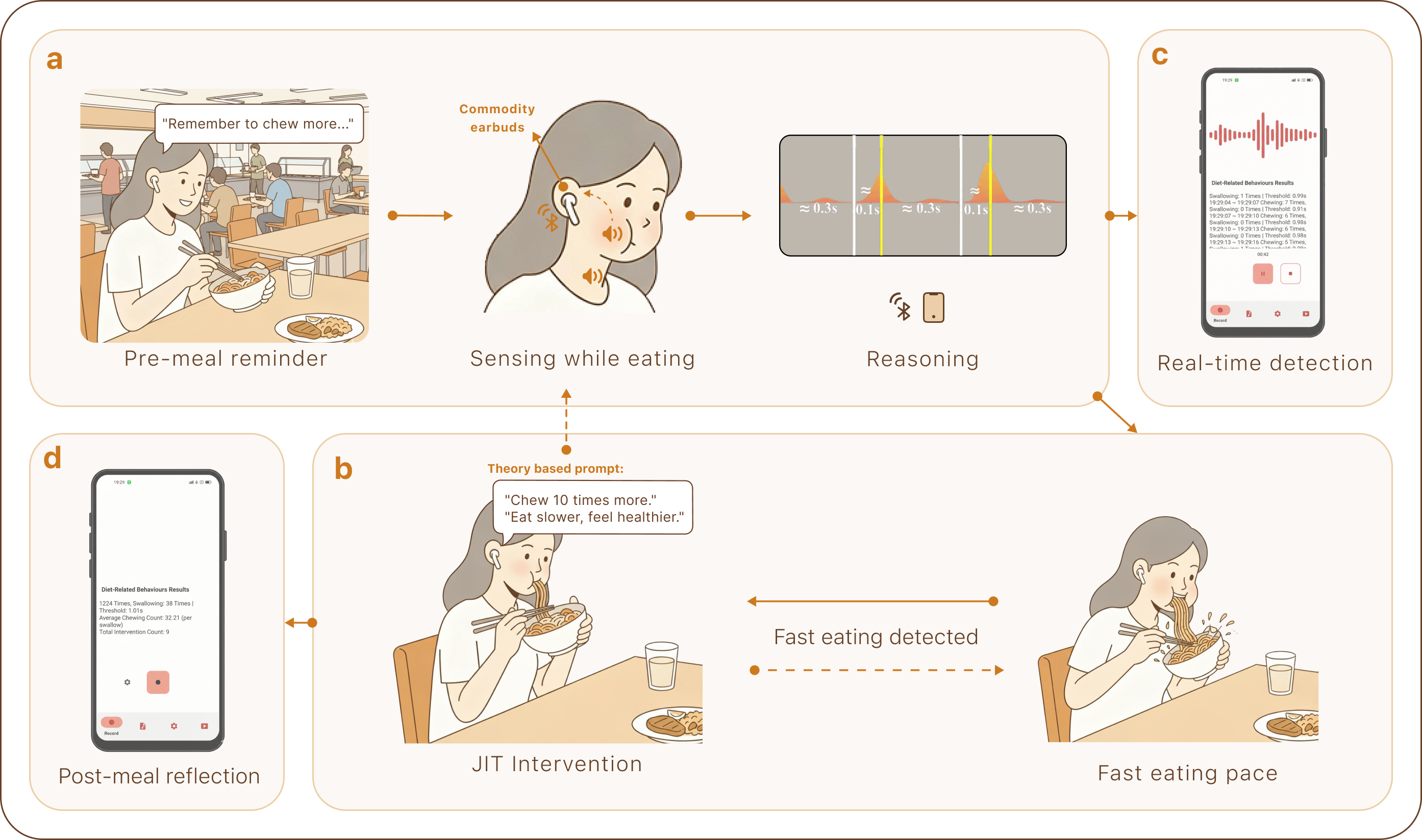}
    \caption{Earinter is an earbud-based closed-loop system for monitoring and regulating eating pace in daily meals. \textbf{(a)} It has a pre-meal reminder, senses ingestive signals while eating in the wild, and performs real-time reasoning to estimate pace. \textbf{(b)} When a rapid eating pace is detected, it delivers theory-based audio prompts, subject to a cooldown interval. \textbf{(c)} It shows real-time ingestive events and pace estimates. \textbf{(d)} After the meal, it summarizes pace-related statistics to support reflection.}
    \label{fig:teaser}
\end{figure}

Rapid eating is common yet difficult to regulate in situ, partly because people seldom notice pace changes and sustained self-monitoring is effortful. We present Earinter, a commodity-earbud-based closed-loop system that integrates in-the-wild sensing, real-time reasoning, and theory-grounded just-in-time (JIT) intervention to regulate eating pace during daily meals. Earinter repurposes the earbud's bone-conduction voice sensor to capture chewing-related vibrations and estimate eating pace as chews per swallow (CPS) for on-device inference. With data collected equally across \modify{sessions in quiet and noisy environments}, Earinter achieves reliable chewing detection (F1 = 0.97) and accurate eating pace estimation (MAE: 0.18 $\pm$ 0.13 chews/min, 3.65 $\pm$ 3.86 chews/swallow), enabling robust tracking for closed-loop use. Guided by Dual Systems Theory and refined through two Wizard-of-Oz pilots, Earinter adopts a user-friendly design for JIT intervention content and delivery policy in daily meals. In a 13-day within-subject field study (N=14), the closed-loop system significantly increased CPS and \modify{slowed meal-level eating pace}, with statistical signs of carryover on retention-probe days \modify{and generally favorable comfort and usability}. These effects should be interpreted as changes in eating-pace regulation rather than evidence of weight loss. Our findings highlight how single-modality commodity earables can support practical, theory-driven closed-loop JIT interventions for regulating eating pace in the wild.

\end{abstract}



\maketitle
\section{INTRODUCTION}
\label{intro}


Eating is a foundational daily activity with downstream consequences for health, quality of life, and well-being~\cite{jovanov2014sensors, mesas2012selected}. 
While prior work and public attention largely emphasize what people eat (e.g., diet quality~\cite{sun2020postcard, tang2015introducing} and nutritional balance~\cite{alshurafa2015recognition, faltaous2021wisdom, bedri2022fitnibble}), how people eat during a meal is equally consequential. We define \textbf{eating pace} as the temporal pattern of within-meal food intake, capturing how quickly and steadily a person progresses through eating. However, it is often overlooked in everyday life~\cite{wilkinson2016keeping}.


In everyday meals, regulating eating pace matters because persistent fast eating has been associated with delayed satiety and higher risks of overeating and obesity~\cite{goto2015influence, nicklas2001eating, zhu2013increasing}, increased gastrointestinal burden due to insufficient chewing~\cite{martins2024association, li2013habitual}, and reduced awareness of the eating experience~\cite{robinson2013eating}. Yet it is difficult to regulate in situ. People often fail to notice when they speed up~\cite{melanson2023eating}, cannot sustain attention across the meal~\cite{mantzios2025eating}, and face strong habitual stability reinforced by daily and social contexts~\cite{chen2022slnom,goh2021impact, kim2016eating, chen2023factors}. As a result, eating pace becomes a largely automatic behavior with little immediate feedback, allowing fast-eating patterns to persist unchecked. 

Recent advances in ubiquitous and wearable sensing have made it increasingly feasible to detect ingestive behaviors such as chewing and swallowing in the wild and even run online inference during daily meals~\cite{lotfi2020comparison, kamachi2021automatic, chen2021chewpin, zhang2022eatingtrak, shin2022mydj, zhang2022enabling}. In practice, however, most existing work still has limited translation of these behavioral signals into actionable, acceptable, and timely support~\cite{fang2025review}. When interventions are explored, they are often either after the fact (e.g., post-meal visual summaries for reflection)~\cite{faltaous2021wisdom, hori2023masticatory, nakamura2023eat2pic} or workflow-bound in-meal cues (e.g., video playback changes~\cite{chen2025vifeed}, replayed eating sounds~\cite{chen2022slnom}, vibrotactile prompts~\cite{kim2016slowee, hermans2017effect}). A key reason is that eating pace is dynamic, context-dependent, and habit-driven -- it fluctuates within meals and varies across people, foods, and social settings -- making post-meal feedback too late and fixed rules brittle. What is still missing is an iterative closed-loop mechanism that continuously senses and infers momentary eating states and maps them to intervention timing and dose (when to intervene, how often, and for how long), enabling support that remains effective and acceptable in everyday meals.


\modify{To address this gap, we present Earinter, a commodity-earbud-based closed-loop system for eating-pace regulation during daily meals. We treat sensing and intervention as components of a single closed loop rather than as separate stages. We choose earables not only for sensing, but also because they co-locate sensing and private, eyes-free feedback on an already-adopted form factor, reducing setup friction and limiting competition with visual attention during meals. Earinter repurposes the bone-conduction voice-pickup sensor already embedded in commodity earbuds, a design pattern originally intended for call enhancement and now common in mainstream devices. This practical choice can capture chewing-related vibrations and infer chews per swallow (CPS) as an actionable indicator of oral processing intensity~\cite{sanchez2013relationship, goh2021increased}. Across data collected in both \modify{quiet and noisy} settings, Earinter achieved reliable chewing detection (F1 = 0.97) and low error in pace-relevant estimation (MAE: 0.18 $\pm$ 0.13 chews/min; 3.65 $\pm$ 3.86 chews/swallow). Then Earinter uses these real-time estimates to trigger just-in-time support. To make support practical in everyday eating, we further design a theory-grounded intervention mechanism guided by Dual Systems Theory and refine content and delivery policy through two Wizard-of-Oz pilots. We then evaluated the full system in a 13-day within-subject field study (N=14), where Earinter significantly increased CPS and reduced food-consumption speed, with statistical signs of carryover on retention-probe days and overall comfort and usability. Together, these results show the promise of commodity earables as a practical platform for theory-driven, closed-loop JIT intervention for eating-pace regulation in the wild. In this paper, we make the following contributions:}


\begin{itemize}
    \item We present Earinter, a commodity-earbud-based closed-loop system that regulates eating pace in daily meals by continuously sensing and reasoning about ongoing ingestion, delivering JIT auditory interventions in situation and a post-meal summary report.
    \item We develop a theory-grounded, psychologically informed JIT intervention mechanism. A Wizard-of-Oz pilot study refines the final prompt set and calibrates delivery parameters, including frequency and sequencing, to \modify{preserve comfort while maintaining perceived usefulness}.
    \item We evaluate Earinter in a 13-day within-subject field study with baseline, control, and experimental phases, showing significant improvements in eating pace and statistically early signs of carryover with \modify{favorable comfort and usability.}
\end{itemize}

\section{BACKGROUND AND RELATED WORK}


\subsection{Eating Pace as a Dynamic In-Meal Behavior}


Eating pace is a continuous, within-meal behavior that reflects the moment-to-moment rhythm of ingestion. It is closely related to mindful eating in that both emphasize awareness of the eating process and attention to bodily cues \cite{sonoki2013effects}. Persistent fast eating pace has been linked to a range of adverse outcomes, including greater risks of overeating and obesity \cite{goto2015influence, ohkuma2015association, otsuka2006eating}, higher cardiometabolic burden \cite{yamaji2017slow, radzevivciene2013fast}, and even acute safety hazards \cite{kumar2008fast}. Mechanistically, fast eating pace often entails insufficient mastication and hurried swallowing, which can increase gastrointestinal load and reduce the effectiveness of digestive processing \cite{li2013habitual}. Meanwhile, chewing-related physiological feedback (e.g., appetite and glucose regulation) may lag behind rapid intake \cite{zhu2013increasing, read1986swallowing}, making satiety signals arrive too late to shape ongoing behavior and thereby encouraging excessive energy intake \cite{alsalim2019insulin}. In everyday meals, eating pace is shaped by time pressure, situational norms, and social contexts, and fast eating is often \delete{socially }acceptable \cite{kim2016eating, chen2023factors}. Eating is also often treated as a secondary companion activity during leisure-time, and people frequently engage in screen media or other distractions during meals \cite{chen2025vifeed}. This makes it difficult to sustain attention to eating pace throughout a meal, especially in the absence of immediate corrective feedback. Moreover, eating pace tends to exhibit habitual stability across contexts, which weakens the effectiveness of intention-based regulation strategies  \cite{chen2022slnom,goh2021impact}.

Building on prior work, we operationalize eating pace through oral processing intensity, quantified as chews per swallow (CPS). CPS is defined as the number of chewing cycles between two consecutive swallowing events \cite{sanchez2013relationship, goh2021increased}. As an intuitive behavioral indicator, CPS is more actionable than other chewing-related metrics, which are regulated by the rhythm generator in the central nervous system and therefore tend to vary only modestly in response to external influences \cite{lund2006generation, aoshima2025greater,  morquette2012generation}. In contrast, CPS can be adjusted through simple, in-the-moment strategies such as extending mastication before swallowing. Recent work increasingly adopts CPS to characterize eating pace and examine its links to health outcomes \cite{zhu2013increasing, borvornparadorn2019increased, goh2021increased}. In this paper, CPS serves as the primary continuous indicator for monitoring and regulating eating pace during daily meals.

\subsection{JIT Intervention Design for Ingestion Behavior Change}
\label{chap:JIT_theory}


Just-in-time (JIT) interventions refer to methods that deliver support when an individual is most likely to benefit from and accept it \cite{nahum2016just}. Such moments are typically identified through real-time assessments of the user's behavior, physiology, or context \cite{sarker2014assessing, nahum2016just, choi2019multi}. Beyond timely delivery, effective JIT design also requires a principled account of why the target behavior occurs and is hard to change. Accordingly, prior researches in behavior change emphasize grounding JIT intervention in psychological theories to improve acceptability and increase the likelihood of sustained habit change \cite{webb2010using, michie2008theory}. In digital well-being, Dual Systems Theory \cite{kahneman2011thinking} frames problematic smartphone use as a conflict between habitual and deliberative control processes \cite{pinder2018digital}. It motivates interventions that adapt to in-the-moment triggers such as fatigue, emotions, and context \cite{wu2024mindshift}. In medication adherence, Self-efficacy Theory \cite{bandura1997self} motivates feedback that reinforces success and supports recovery after lapses \cite{lee2014real}. 

In eating-related interventions, recent systems have increasingly incorporated theories to strengthen long-term change in dietary behaviors. Self-efficacy Theory has motivated designs that provide positive reinforcement, small achievable actions, or peer support to encourage healthier food choice routines \cite{rabbi2015mybehavior, kim2017mobile}. Goal-oriented theories further strengthen goal-directed behavior in complementary ways -- Goal-setting Theory operationalizes distal goals into proximal subgoals with progress feedback \cite{patra2024personal}; Control Theory frames self-regulation as discrepancy monitoring and closed-loop correction \cite{hietbrink2023digital}; and Goal-framing Theory increases goal salience by shaping how rationales and benefits are communicated \cite{achananuparp2018eat, de2022effectiveness}. However, these theory-driven work focuses on behavior change outside the eating episode, and theory is often used for after-the-fact interpretation rather than for in-the-moment intervention design \cite{fang2025review}. When considering JIT interventions during eating episodes, existing approaches often rely on interactive artifacts to deliver timely cues, such as game-based installations \cite{ganesh2014foodworks, kadomura2014persuasive}, interactive robots \cite{robinson2021humanoid, xie2023chibo}, or transformable tableware \cite{chen2022sspoon, khot2020swan}. Other work probes behavioral effects by manipulating elements within a pre-defined routine, such as adjusting video playback speed \cite{chen2025vifeed} or replaying eating-related sounds \cite{chen2022slnom}. 


\modify{However, to our knowledge, no prior system has grounded both in-meal intervention content and explicitly designed delivery policies (e.g., when to intervene, how often, and for how long) in a shared theory-driven rationale for habit-like behavior that fluctuates dynamically within meals~\cite{faltaous2021wisdom, hori2023masticatory, nakamura2023eat2pic, chen2025vifeed, chen2022slnom, kim2016slowee, hermans2017effect, fang2025review}. 
Motivated by this gap, Earinter uses theory not only to shape what support is delivered, but also to inform when support is triggered and how intervention dose is regulated during the meal. We use theory to characterize the habitual behavior, guide the design of prompt content, and inform an adaptation logic that responds to real-time eating pace. This framing lays the foundation for more effective and personalized closed-loop support for eating-pace regulation.}
\label{2_2}

\subsection{Closed-loop System}


The closed-loop system is inspired by the Personal Informatics Framework \cite{li2010stage}, mapping collection to sensing, integration to reasoning, and the reflection-action sequence to intervention \cite{fang2025review}. It's an iterative cycle that uses continuous sensing and reasoning to deliver context-relevant support and update subsequent actions based on users' moment-to-moment responses. 
This paradigm has increasingly been adopted in ingestion health. However, systems for ingestion still face persistent challenges in achieving end-to-end closed-loop support. 
First, intervention outputs are often limited to lightweight cue modalities, including audio \cite{chen2022slnom} and visual cues \cite{nakamura2023eat2pic,kim2016slowee, wilkinson2016keeping}, whatever sensing pathways and modalities are. Consequently, sensing and intervention remain weakly integrated, with limited coupling between sensing outputs and feedback pathways. Second, many pipelines depend on multiple devices to realize the full loop, which increases setup complexity and users' cognitive burden. Third, systems often prioritize recognition accuracy while treating intervention as a downstream components rather than a theory-informed design problem. As a result, adaptation logic for when to intervene, how frequently, and how to respond to short-term behavioral changes remains under-specified.

For eating pace in particular, much prior work emphasizes sensing and reasoning, shifting from self-report to wearable based monitoring \cite{hossain2025systematic}. When interventions are included, the lack of theory grounding often leads to limited design choices. Some systems mainly support post-meal reflection with summaries \cite{scisco2011slowing, faltaous2021wisdom}, while others use simple but information-sparse in-meal cues such as vibration \cite{hermans2017effect, kim2016slowee}. More expressive in-meal interfaces, including real-time visualizations \cite{nakamura2023eat2pic, kim2018animated, Buntsma2023}  or transformative utensils \cite{chen2022sspoon, zhang2020exploring}, can increase information load and potentially compete with food awareness. Collectively, these patterns motivate closed-loop eating-pace systems that tightly integrate sensing, reasoning, and theory-informed intervention, while specifying adaptation policies that balance effectiveness with practical acceptability in daily meals.

\section{SYSTEM OVERVIEW}
\label{system_overview}


\begin{figure}[htbp]
    \centering
    \includegraphics[width=.9\linewidth]{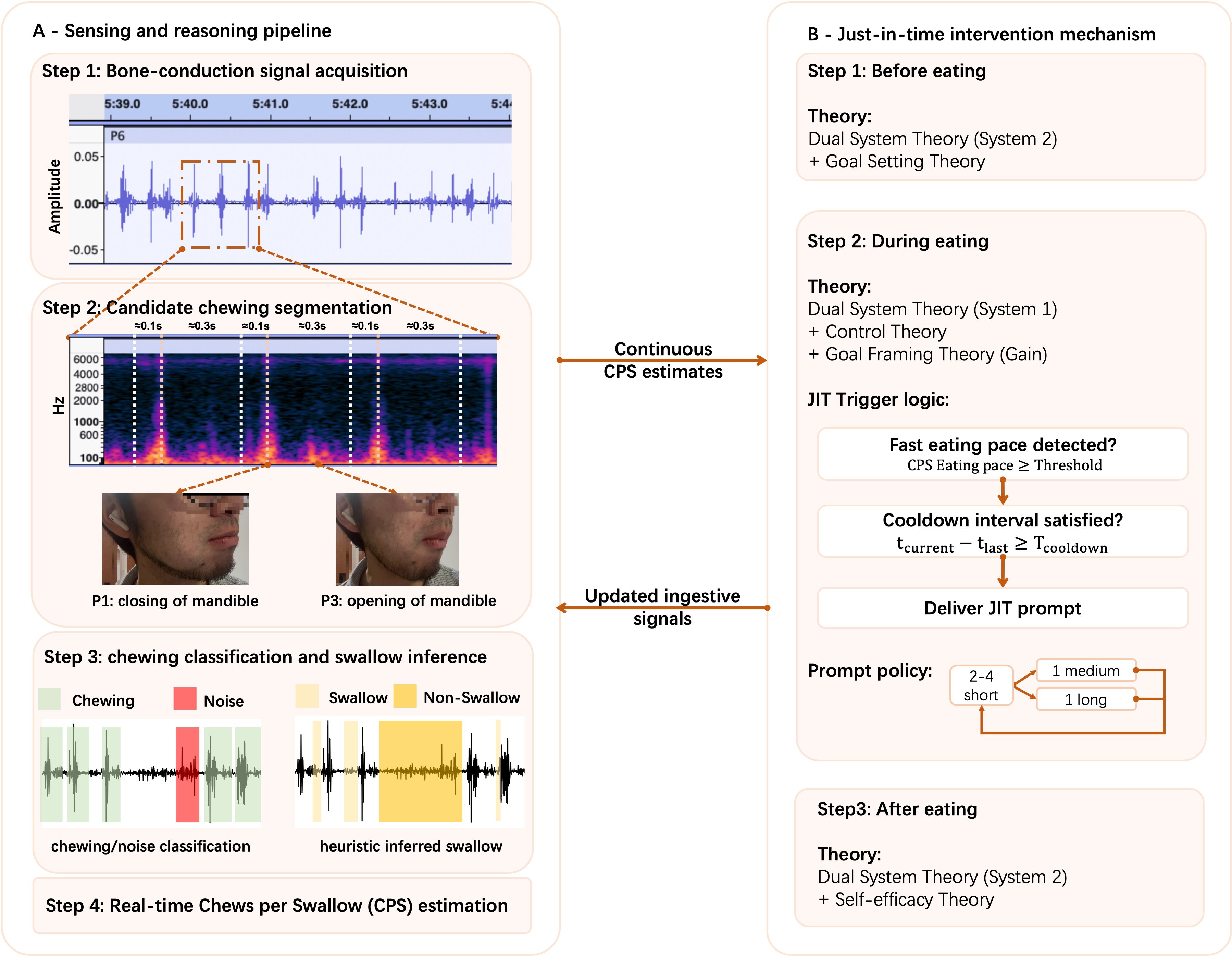}
    \caption{\modify{Closed-loop system overview of Earinter for eating-pace regulation. (a) Sensing and reasoning pipeline. The commodity earbud captures bone-conducted ingestive signals during eating. Earinter first acquires the raw signal, segments candidate chewing events, classifies chewing versus noise, and infers swallow boundaries heuristically, producing real-time chews-per-swallow (CPS) estimates. (b) Just-in-time intervention mechanism. Continuous CPS estimates drive theory-grounded intervention across the eating episode, including pre-meal goal activation, in-meal JIT prompting, and post-meal reflection. During eating, prompts are triggered when fast eating pace is detected and the cooldown interval is satisfied, with diversified prompt lengths used to reduce burden. The resulting behavioral response generates updated ingestive signals, which are continuously re-sensed by Earinter to complete the closed loop.}}
    \label{fig:system_overview}
\end{figure}

\add{Before detailing the architecture, we clarify Earinter's position relative to representative strands of prior eating-related systems. Reflection-oriented systems primarily support awareness after the eating episode \cite{faltaous2021wisdom, hori2023masticatory}; fixed in-meal cueing systems deliver prompts during eating but often use predefined delivery or focus on workflow-bound cues~\cite{kim2016slowee, hermans2017effect, chen2022slnom, chen2025vifeed}; and sensing-centric ingestion systems mainly advance recognition and monitoring, with intervention typically treated as a downstream step \cite{chen2021chewpin, lotfi2020comparison, kamachi2021automatic, zhang2022eatingtrak, shin2022mydj, bedri2017earbit,bi2018auracle}. Earinter extends these strands by integrating real-time CPS estimation from commodity-earbud sensing and a theory-grounded JIT intervention mechanism within a closed-loop system for daily meals.}
\add{We use commodity earables as the system substrate because they can unify practical sensing and private, in-ear intervention within a single practical form factor.} As shown in Fig.~\ref{fig:system_overview}, Earinter is a closed-loop system using commodity earbuds, which comprises three stages: sensing, reasoning, and intervention.

\paragraph{Sensing.}
Earinter repurposes the bone-conduction sensor embedded in a commercial earbud (\textbf{Honor Earbuds 3 Pro}\footnote{\blueurl{https://www.honor.com/global/audio/honor-earbuds-3-pro/}}) to capture body-conducted chewing and swallowing signals with reduced sensitivity to ambient noise. We describe the sensing setup and signal acquisition in Section~\ref{sec:sensing_setup}.

\paragraph{Reasoning.}
\modify{
Signals are processed on an Android smartphone (\textbf{OPPO Reno5 PRO}\footnote{\blueurl{https://www.oppo.com/en/smartphones/series-reno/reno5-pro-5g/}}) for real-time inference. To validate the sensing pipeline before field deployment, we first describe a separate annotated sensing dataset collected across complementary \modify{quiet and noisy} contexts in Section~\ref{sec:data_collection}, followed by the detection model in Section~\ref{sec:sensing_pipeline} and model validation in Section~\ref{chap:model_evaluation}. 
}

\paragraph{Intervention.}
When inferred CPS indicates fast eating, Earinter delivers JIT audio prompts through the earbud to encourage a slower pace. The intervention framework is grounded in Dual System Theory (Section \ref{chap:jit_dual}), with prompt content informed by relevant physiological theories (Section \ref{chap:jit_content}), and the delivery policy including logic and frequency refined for practical use in daily meals (Section~\ref{chap:jit_logic}).

\section{REAL-TIME EATING PACE SENSING STRATEGY ON EARABLES}
\label{chap:algorithm}



\begin{figure}[h]
    \centering
    \includegraphics[width=.9\linewidth]{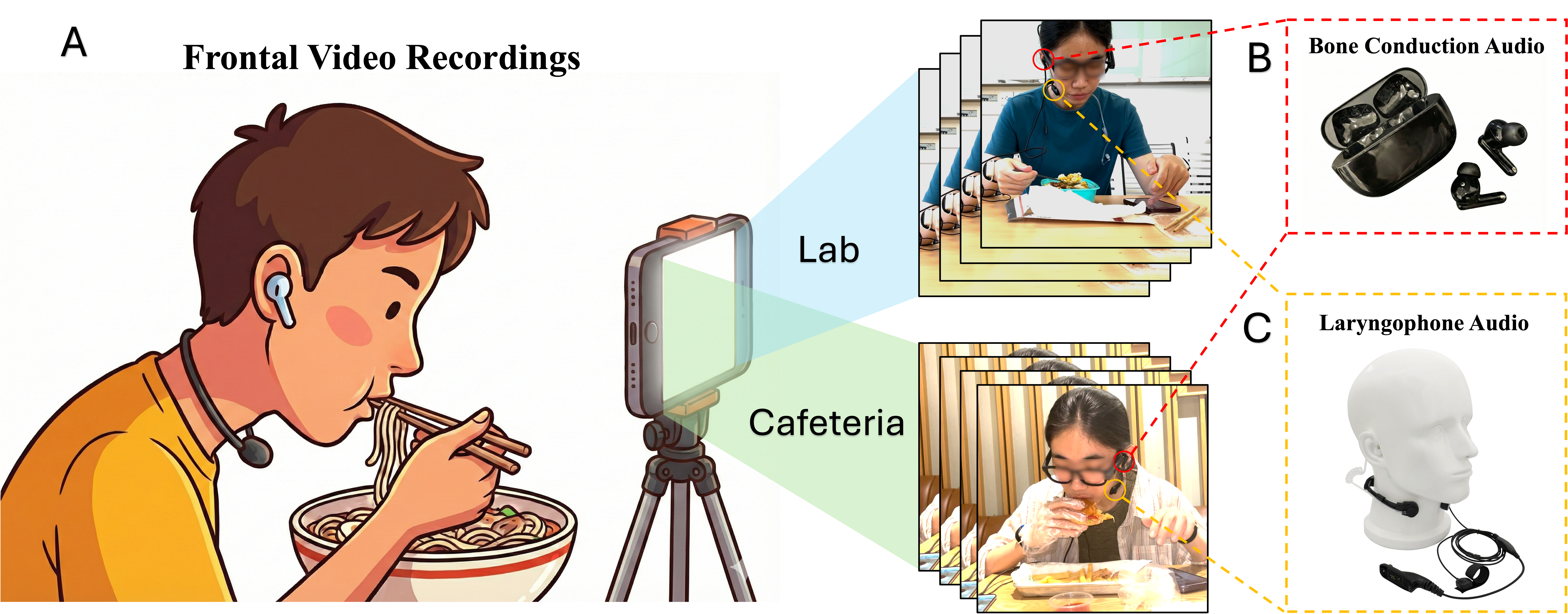}
    \caption{\modify{Sensing-data collection setup and cross-context recording protocol used to validate the ingestion sensing pipeline.} Eating behaviors were recorded across two environments (lab and cafeteria) using: \textbf{(A)} an eye-level iPhone for video ground truth, \textbf{(B)} integrated bone-conduction sensors in Honor Earbuds 3 Pro for capturing ear-canal vibrations, and \textbf{(C)} a laryngophone serving as a high-fidelity reference for precise chewing and swallowing annotation.}
    \label{fig:data_collection}
\end{figure}

\subsection{Data Collection}

\subsubsection{Hardware Setup}
\label{sec:sensing_setup}

We designed a data collection platform that measures eating pace, which we define as the number of chewing cycles between two consecutive swallows. \textbf{Honor Earbuds 3 Pro} leverages integrated bone-conduction (BC) sensors to capture chewing-induced vibrations within the ear canal. The devices were provided through a collaborative research agreement with Honor Device Co., Ltd. To establish a reliable reference for identifying chewing and swallowing events, a \textbf{DZMIHKS6550 throat microphone}~\footnote{\blueurl{https://www.amazon.com/Earpiece-Acoustic-Compatible-MMotorola-XiRP8268/dp/B016BI0CD4/}} was concurrently employed alongside an \textbf{iPhone 12 Pro}~\footnote{\blueurl{https://www.amazon.com/Apple-iPhone-128GB-Pacific-Blue/dp/B08PMYLKVF?th=1}} serving as a rear-facing camera. Audio was sampled at 16 kHz with 16-bit resolution, while video was recorded at 1080p with a frame rate of 30~fps. Synchronizing these multi-modal streams enabled precise temporal calibration to ensure the accurate determination of ingestion events.

\subsubsection{Data Collection Protocol} 
\label{sec:data_collection}

\add{To validate the sensing pipeline before the main field deployment, we collected a separate annotated sensing dataset across two complementary meal-recording settings. Food choice was unrestricted in both settings, so the main contrast was the recording environment. The quiet-environment sessions took place in a lab conference room to approximate quiet individual eating scenarios, whereas the noisy cafeteria sessions introduced realistic variation in ambient sound and surroundings while keeping the sensing configuration comparable across settings.}

This study was approved by our Institutional Review Board. Participants were recruited via word-of-mouth from our campus. Informed written consent was obtained from the human subjects participating in the study before any study procedures began. We recruited 18 participants (9 females, 9 males) aged 20 to 29 years ($M = 23.83$, $SD = 2.66$). All participants were healthy adults with no history of gastrointestinal or metabolic diseases. No participants reported any pain or discomfort while chewing or swallowing.

As shown in Fig.~\ref{fig:data_collection}, each participant completed sessions in two settings: \modify{a quiet lab conference room and a noisy campus cafeteria} chosen by the participant. To ensure ecological validity, sessions were conducted during regular meal times, and participants were free to select their food. Before each session, researchers assisted participants with device wearing. Participants wore earbuds in both ears to capture bone-conduction audio, while a throat microphone was worn on the neck to provide reference data. The eye-level iPhone 12 Pro recorded the meal discreetly to avoid interfering with natural behavior. To synchronize data streams, participants double-tapped the earbuds on the table at the start of the session. Participants were encouraged to eat at their natural pace and avoid any intentional control over their chewing or swallowing to ensure the data reflected their genuine habits.

The dataset comprises 10.07 total hours of data, including 5.65 hours from \modify{quiet} and 4.42 hours from \modify{noisy} settings. Bone conduction audio was manually annotated using Praat~\cite{Boersma2026Praat} and ELAN~\cite{ELAN2025} as annotation tools. Through the annotation process with video and throat microphone audio as reference, we identified 24,521 and 21,061 chewing events, alongside 1,114 and 1,036 swallows for the sessions \modify{in quiet and noisy environments}, respectively. Detailed statistical results are shown in Table~\ref{tab:eating-dataset-appendix}.

\add{To characterize food variation in this dataset, we further conducted a coarse meal-level texture annotation. Because meals often contained multiple components, categories were not mutually exclusive: all 36 meals included soft components, 23 included crunchy components, and 18 were annotated as clearly mixed-texture meals. Examples ranged from softer noodle-based dishes to crunchier meals such as fried chicken and leafy salads. These annotations indicate that the dataset covered heterogeneous everyday meals rather than a narrowly controlled single-food benchmark, better reflecting the practical meal contexts targeted by Earinter for daily eating-pace regulation.}

\subsection{Eating Pace Sensing Pipeline}
\label{sec:sensing_pipeline}

\begin{figure}[h]
    \centering
    \includegraphics[width=1.0\linewidth]{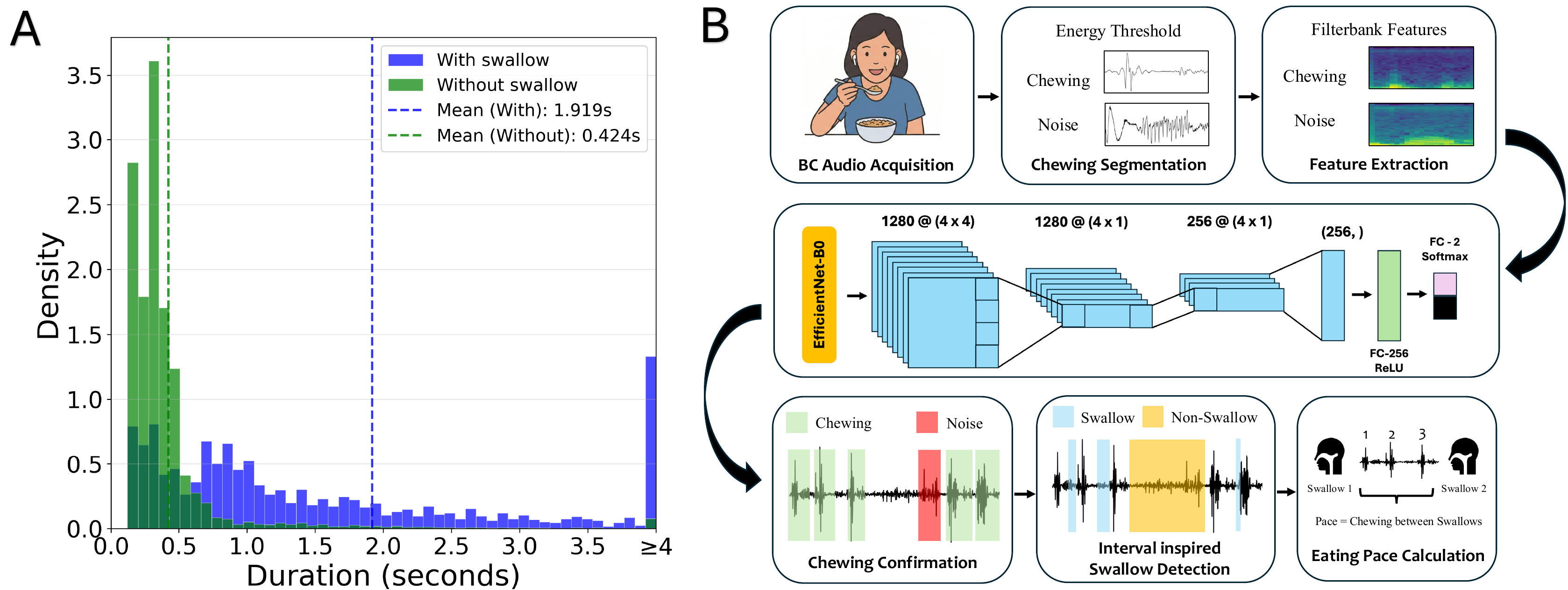}
    \caption{Swallow detection foundation and the eating pace sensing pipeline. \textbf{(A)} Probability density distribution of durations between consecutive chewing cycles. The distinct temporal separation between intervals with swallows ($\mu = 1.919s$) and without swallows ($\mu = 0.424s$) validates the interval-inspired hypothesis for swallow detection. \textbf{(B)} The proposed sensing pipeline, which integrates \modify{candidate} chewing segmentation, deep learning-based chewing confirmation, interval-inspired swallow detection to estimate the eating pace.}
    \label{fig:sensing_pipeline}
\end{figure}

Fig.~\ref{fig:sensing_pipeline} depicts the sensing pipeline for eating pace monitoring on earables. We detail each stage below.

\subsubsection{Audio Pre-processing and Candidate Chewing Segmentation}
\label{sec:chewing_segmentation}
\modify{Chewing is modeled as a four-stage mandibular motion cycle: food compression during mandible closing (P1), a brief pause (P2), mandible opening (P3), and another pause before the next cycle (P4)~\cite{zhang2012mastication}.} \add{P2 and P4 barely generate acoustic energy, and therefore they are not used for chewing segmentation.} As shown in Fig.~\ref{fig:system_overview}, P1 represents the high-energy phase of closing the lower chin to crush food \add{compared to P3}. We first implement a real-time state machine to extract candidates by partitioning audio into 50~ms frames and applying a 40~dB threshold. To align with physiological patterns, valid segments are limited to a duration between 100~ms and 400~ms with a maximum silence tolerance of 150~ms. These parameters were empirically determined to ensure that all chewing events within the dataset were captured without omission. This lightweight screening optimizes system efficiency by filtering non-eating noise and reducing the computational load for downstream deep-learning classification.

\subsubsection{Fine-grained Chewing Detection} 
\label{sec:fine_grained_chewing_detection}

To filter out motion artifacts introduced during the initial screening, we implement a binary classification model. The dataset is partitioned into 14 participants (77.8\%) for training, 2 for validation (11.1\%), and 2 for testing (11.1\%), which  facilitates cross-user evaluation. Each candidate segment was normalized to a range between $[-1, 1]$ and padded to a uniform duration of 400 ms.

Inspired by the EarVAS-Net framework~\cite{zhang2024earsavas}, the windowed audio is transformed into log-scaled Mel filter bank features of size 38 x 128 using a 25 ms Hanning window with a 10 ms stride. We adapt the EarVAS audio sub-module for single-channel processing by utilizing EfficientNet-B0 as the feature extraction backbone. This architecture incorporates Mean Pooling along the frequency axis, a $1\times1$ convolutional layer, and subsequent Mean Pooling along the time axis. This streamlined configuration maintains the robust feature extraction of EfficientNet-B0 while optimizing computational efficiency for real-time binary classification. \delete{on earable devices.} To address the inherent class imbalance between genuine chewing events and non-eating artifacts, we adopted Focal Loss~\cite{lin2017focal} as the loss function. The model was trained on an NVIDIA GeForce RTX 4090 GPU using a batch size of 128 and a learning rate of $1e^{-5}$.

\subsubsection{Heuristic Estimation of Eating Pace} 
\label{sec:heuristic}
Eating pace was defined as the number of chewing events between two consecutive swallows. However, the commercial earbuds that we used often fail to capture swallowing events due to their subtle acoustic signatures. To \modify{estimate} eating pace, we propose a heuristic algorithm based on the inter-chew interval. \add{In the chewing cycle, P4 corresponds to the pause after mandibular opening and before the next chewing cycle. When swallowing occurs, this pause is often prolonged, creating a longer interval between consecutive chewing events.} Specifically, our statistical analysis indicates that swallowing events are primarily localized within the prolonged intervals between chewing events, as shown in Fig.~\ref{fig:sensing_pipeline}A. This observation aligns with the natural eating experience. Based on the observations, once the chewing detection algorithm \modify{locates} chewing events, the presence of swallowing within the remaining temporal gaps is determined by the following heuristic algorithm:

\begin{equation}
isSwallowing = \begin{cases} 
yes, & \text{if } d > 0.8 \text{ s} \\ 
yes, & \text{if } d > 1.5 \times \bar{d}_{\text{chew}} \\ 
no, & \text{otherwise} 
\end{cases}
\end{equation}

where $d$ represents the duration of the current interval between consecutive chewing events, while $\bar{d}_{\text{chew}}$ denotes the average duration of all chewing intervals recorded since the most recent swallow. The absolute threshold (0.8~s) captures prolonged intervals likely containing a swallow while limiting false positives. To account for individual- and food-dependent chewing tempo, we also apply an empirically chosen adaptive criterion (1.5 $\times \bar{d}_{\text{chew}}$), enabling swallow inference and CPS estimation without explicit swallow acoustics.

\subsection{Evaluation and On-device Deployment}
\label{chap:model_evaluation}

\subsubsection{Evaluation of Chewing Detection}

Since the system performs binary classification on extracted candidate segments, we assess the performance of the chewing detection module using four standard metrics. On the test dataset, the model achieves an Accuracy of 96.86\%, Precision of 99.02\%, Recall of 95.05\% and an F1-score of 0.97. These results demonstrate the high reliability of the pipeline in distinguishing genuine chewing events from artifacts.

\subsubsection{Evaluation of Eating Pace Estimation}

To validate the temporal precision of the heuristic algorithm, we analyze eating pace through two complementary dimensions. At a macro-level, we measure the mean estimation error of the chewing frequency per minute to characterize the overall velocity of intake. In all experimental environments, the system achieves a mean absolute error (MAE) of $0.18 \pm 0.13$.

At a more granular level, we assess the mean estimation error of the number of chews per swallow, which serves to validate the effectiveness of heuristic segmentation in accurately demarcating discrete masticatory sequences between inferred swallowing events. For this fine-grained metric, the system achieves an MAE of $3.65 \pm 3.86$. In summary, these results demonstrate that the proposed system provides \modify{adequate} monitoring of both long-term eating trends and detailed masticatory behavior.

\subsubsection{On-device Inference and Performance} 
\label{sec:on_device_performance}
To enable the intervention of the proposed system, we deployed the sensing framework on an \textbf{Oppo Reno5 Pro}. We developed a dedicated Android application to facilitate real-time interaction and data visualization for the user. To ensure efficient execution on the mobile devices, the deep learning classification model was converted to the ONNX format. The entire pipeline processes incoming acoustic data from the earable device using a 3-second sliding window. 

\add{We measured power consumption of our system using Android batterystats during one hour of screen-off background execution. Power consumption is approximately 1.84\% battery capacity per hour on Oppo Reno5 Pro with a 4350~mAh battery. According to the data collected, eating-episode duration ranged from 5.25 to 29.02 minutes. This suggests that a typical meal-length would consume approximately 0.16\%-0.89\% of the phone battery. As typical smartphone applications like music players consumes about 2\% per hour~\cite{Hu2024Detecting}, our algorithms bring limited battery burden for daily meal-time deployment.}

\add{To guage system performance, we measure CPU load and latency by running a specific code segment 1000 times.} The system processes a 3-second audio stream in just 57.64 $\pm$ 24.89~ms, demonstrating its high efficiency for real-time eating pace estimation and just-in-time intervention.
\section{PILOT STUDY FOR INTERVENTION DESIGN}
\label{chap:pilot}
Given that Earinter senses ingestive signals using commodity earbuds, we deliver intervention through auditory prompts to support modality-consistent closed-loop system. This design locates sensing and intervention on the same wearable channel, which reduces setup complexity and cognitive burden in everyday use. Auditory prompts are also hands-free and eyes-free, which better aligns with the attentional demands and social nature of daily meals. Moreover, language is the most common persuasion \modify{mechanism} in behavior change methods \cite{rapp2022aristotle, simons1976persuasion}. Concise reminders grounded in facts or goals can effectively support the in-the-moment self-regulation \cite{morris2018towards, kaptein2012adaptive}. 
Accordingly, we implement an earbud-based JIT audio intervention grounded in Dual System Theory. Building on this theory-driven pipeline, we further organize in-meal prompts using additional psychological theories, and conduct two Wizard-of-Oz (WoZ) pilot studies to refine (PS1) prompt content and (PS2) prompting frequency for practical, in-the-wild use.

\subsection{Dual System Theory for Intervention Design}
\label{chap:jit_dual}

As described in Section~\ref{chap:JIT_theory}, drawing on prior work on intervening in habitual behaviors \cite{wu2024mindshift, chen2025vifeed}, we ground our intervention design in Dual System Theory \cite{kahneman2011thinking}. Dual System Theory describes the structure of cognitive processing: fast, automatic/implicit processes (System 1) versus slow, controlled/deliberative processes (System 2).  We conceptualize rapid eating pace as a largely automatic routine driven by System~1, whereas regulating pace requires re-emphasizing System~2 processes, such as attention re-allocation \cite{chen2025vifeed} and self-control \cite{wu2024mindshift}.

Guided by Dual System Theory, we structured prompt content across the full eating episode to support both goal activation (System~2) and in-the-moment disruption of habitual rapid eating (System~1).

\begin{itemize}
    \item \textbf{Before eating (System~2).} We generated prompts informed by Goal Setting Theory \cite{locke1968toward} to make the target behavior explicit and memorable. These prompts emphasize specifying actionable targets to initiate deliberate self-regulation, aligning with System~2 processing.
    \item \textbf{During eating (System~1).} We deliver brief prompts as intuitive cues that shape automatic pacing with minimal cognitive demand. We treat these prompts as System~1-oriented micro-nudges. They rely on simple, immediate signals rather than analytical reasoning or rule-based feedback, and can be acted upon without sustained deliberation.
    \item \textbf{After eating (System~2).} Drawing on Self-efficacy Theory \cite{bandura1997self}, we provide a brief summary of pace-related statistics and supportive reflection cues to strengthen confidence in behavior change. This component supports sustaining behavior change beyond immediate prompts and thus constitutes part of System~2.
\end{itemize}

After defining this pipeline, we further drew on additional psychological theories summarized in prior behavior change syntheses \cite{webb2010using, kwasnicka2016theoretical} to design candidate in-meal prompt families. These families vary in informational specificity, affective tone, and theoretical intent. We also incorporated a commonly cited chewing heuristic (approximately 25 chews per mouthful) as an intuitive reference in prompt wording \cite{alexander1998news, sonoki2013effects}. The full set of theory-grounded intervention content is summarized in Table~\ref{tab:intervention_content}.

\subsection{PS1: Intervention Content}

\label{chap:jit_content}

\begin{table}[htbp]
    \centering
    \caption{Overview of the theory-grounded prompting pipeline and candidate in-meal prompt families \textit{before} the WoZ refinement. We organize intervention content across three stages: pre-meal and post-meal draw on System~2 processes with additional psychological theories (Goal Setting Theory and Self-efficacy Theory), while in-meal prompts include System~1-compatible micro-nudges, non-verbal cues, and theory-informed messages (Control Theory, Goal Framing Theory).}
    \resizebox{.95\textwidth}{!}{
    \begin{tabular}{p{2cm}c c}
    \toprule
    \textbf{Timestamp} & \textbf{{Theory Based}} &  \textbf{Example Content Design}\\
    \hline
    \hline
    Before Eating  & \makecell{Dual System Theory (System~2) \\ + Goal Setting Theory \cite{locke1968toward}} & \makecell{\textit{"Remember to chew more.}  \\ \textit{Aim for at least 25 chews before swallowing."}}\\
    \hline
    \multirow{10}{*}{During Eating}  & {Dual System Theory (System~1)} & \textit{"Slow down." "More chewing." "A bit fast."}\\
    \cdashline{2-3}[1pt/2pt]
      & {None} & Neutral chime cue (\textit{"ding"}) \\
      \cdashline{2-3}[1pt/2pt]
      & {None} & Failure jingle cue \\
      \cdashline{2-3}[1pt/2pt]
     & {Control Theory \cite{carver1982control}} & \textit{"You are xxx chews away from the recommended 25 chews."} \\
     \cdashline{2-3}[1pt/2pt]
      & {Goal Framing Theory (Gain) \cite{tversky1981framing,lindenberg2007normative}} & \makecell{\textit{"Eating more slowly and chewing more can improve digestion} \\ \textit{and increase satiety, helping you reach your weight-loss goal sooner."}} \\
      \cdashline{2-3}[1pt/2pt]
      & {Goal Framing Theory (Loss) \cite{tversky1981framing}} & \makecell{\textit{"Eating too fast and swallowing with insufficient chewing may cause} \\ \textit{digestive discomfort and, contribute to weight gain over time."}} \\
      \cdashline{2-3}[1pt/2pt]
      & {Mixed} & \makecell{A chime is played upon each rapid-pace detection, \\ and a spoken reminder is delivered after repeated detections} \\
    \hline
     After Eating & \makecell{Dual System Theory (System~2) \\ + Self-efficacy Theory \cite{bandura1997self}} & Summary of pace-related statistics \\
     \bottomrule
    \end{tabular}
    }
    \label{tab:intervention_content}
\end{table}


We conducted a Wizard-of-Oz (WoZ) pilot study (N=6) across diverse daily meal contexts: two solo meals, two dyadic meals, and two family co-present meals. Participants also differed in auditory sensitivity.
This allows us to assess perceived usefulness, comfort, and preference across prompt families under individual differences, and to refine wording and content selection prior to deploying the automated intervention.

At the beginning of each session, we explained the study purpose that participants would receive different audio prompts generated from the system during eating \cite{kazdin1982observer}. They are intended to encourage slower eating pace and more chewing. Participants were instructed to eat naturally as usual. During the meal, an experimenter manually triggered prompts in a randomized order, covering the candidate in-meal prompt families shown in Table \ref{tab:intervention_content}. Prompts were delivered through earbuds, with an interval of approximately 60 seconds between consecutive prompts. After all prompts were presented, we sent participants the corresponding text versions and conducted a semi-structured interview. We asked participants to (1) identify prompts they felt were effective versus ineffective, (2) rank most and least preferred prompts, (3) comment on the mixed design, and (4) reflect on perceived annoyance.

All sessions were video-recorded, and interviews were audio-recorded and transcribed. The research team conducted a thematic analysis by coding participants' comments on effectiveness, acceptability, and preferences across prompt families. Codes were iteratively refined through team discussions until consensus was reached, and transcripts were revisited to ensure consistent and accurate coding. 

Overall, participants preferred prompts that were concise, easy to interpret, and low-effort to act on during eating. They noted that longer, information-dense messages could become annoying if delivered frequently (S5\footnote{This is the serial number of the participant in the semi-structured interview for PS1.}, "Words all time takes a lot of effort, and interrupt the entertainment"). So they suggested short prompts for frequent delivery while reserving longer messages for occasional use. Participants also emphasized the importance of content variety to reduce habituation over time. Neutral chime cues alone were sometimes perceived as ambiguous without upfront explanation (S3, "initially thought it was a phone notification"), while failure jingles were viewed as uncomfortable. Moreover, the mixed design was considered acceptable if "an explanation and demonstration of the cue meaning were provided beforehand" (S6).

Based on these findings, we prioritized three in-meal prompt families for subsequent deployment: (i) System~1-oriented micro-nudges, (ii) control-theory-based progress feedback, and (iii) gain-framed rationales. To support practical deployment and reduce repetitiveness, we used GPT-4o to expand each selected family into a prompt library with varied phrasings and multiple lengths, enabling randomized selection at runtime. Generation prompt for GPT-4o and final audio content can be found in Appendix \ref{appendix:inter_content}.





\subsection{PS2: Intervention Frequency}
\label{chap:jit_logic}

Well-crafted prompts can lose effectiveness and become irritating if delivered too often, as shown in PS1. We conducted a second WoZ pilot (N=10) to determine a comfortable maximum prompting frequency, and develop a prompt logic that balances intervention efficacy with user comfort in daily use.

Based on the three categories selected above, we designed three types of prompt lengths: short prompts (approximately 1-2 seconds), medium prompts (2-3 seconds), and relatively long prompts (7 seconds). Then we manipulated the minimum interval between consecutive prompts at three levels: 10 seconds, 20 seconds, and 30 seconds.
These conditions were chosen to span a range from very high frequency (roughly one prompt per bite for a fast eater around 10s \cite{hermans2017effect}) to moderate frequency. Each participant experienced all three interval conditions in a random order. In each condition, five prompts were delivered at the specified interval. To ensure consistency, we fixed the sequence of prompt lengths in each condition to two short prompts, one medium-length prompt, then two short prompts again.  This consistent pattern exposed participants to a mix of brief and slightly longer messages at each frequency. Participants were instructed to eat normally and informed that the system would give messages to encourage slower eating through the earbuds. A 60-second break was provided between each frequency condition to allow participants to rest and minimize carryover effects.

After each condition, participants rated their annoyance with the prompting frequency on a 5-point Likert scale (1 = not annoyed at all, 5 = extremely annoyed). We calculated the scores for each intervention frequency. 
The 10-second interval condition was the least favored (mean = 3.0) and only one participant chose it as prefer. 
The 20-second condition was slightly better (mean = 2.7), and 3 participants preferring it. 
The 30-second minimum interval was the most comfortable by far. It had the lowest annoyance rating (mean = 2.3) and was preferred by 6 out of 10 participants. Participants' feedback clearly indicated that more frequent prompts led to greater annoyance. 
Based on these findings, we set 30 seconds as the minimum interval between prompts in our system design, even if the user remains fast eating pace. 

We also refined the prompt logic from users' feedback to incorporate the content length recommendations from PS1 with the frequency decision above. In the final intervention mechanism, we employ a hybrid prompting strategy to diversify the content and avoid overwhelming users with repetitive messages. The system randomly plays 2-4 short prompts, followed by either a medium-length prompt or a relatively long prompt, with \modify{equal probability}. All these prompts are selected randomly from three families identified in PS1 which are expanded via GPT-4o (Appendix \ref{appendix:inter_content}).





\section{USER STUDY}
\label{user_study}


\modify{Building on the end-to-end pipeline described in Sections~\ref{chap:algorithm}--\ref{chap:pilot}, we evaluated the full Earinter closed-loop system in a within-subject field deployment conducted entirely in participants' everyday in-the-wild eating contexts, including real-time eating-pace inference, theory-based JIT audio prompting with a cooldown policy, and post-meal summaries for reflection.}
We evaluated Earinter against comparison conditions that kept sensing and logging identical but disabled in-meal intervention and post-meal feedback to isolate the contribution of the closed-loop intervention.


\subsection{Recruitment and Study Design}

\subsubsection{Participants}
\label{sec:participants}
We recruited an initial pool of participants who self-reported fast eating pace using snowball sampling through an on-campus communication platform \cite{goodman1961snowball}. All candidates reported being in good general health, having no special dietary habits or dietary restrictions, and having a full set of natural teeth with no pain or discomfort during chewing. \add{Screening was conducted primarily through self-report and online communication, and we did not knowingly enroll participants who self-reported a diagnosed eating disorder or current treatment for an eating-related clinical condition.}

Because self-reports may not accurately reflect eating behavior in everyday settings, we conducted an in-the-wild screening procedure including two meal sessions to confirm eating pace. Specifically, during two free-living meals, we collected ground truth with frontal videos, and assessed eating pace using two commonly adopted criteria: (i) \textit{chews per swallow} $< 20$ \cite{alexander1998news, sonoki2013effects} and (ii) \textit{food consumed per minute} $> 50$ g/min \cite{woodward2020comparison, kissileff1980universal}. Candidates were eligible if they met both criteria in the screening. Based on this screening, we enrolled 14 participants (9 male, 5 female, $M=23.50$, $SD=2.68$) in the main study. These participants completed the full study and received \$200 compensation. \add{The \$200 compensation was provided for completing the full field deployment. Participant effort was distributed across Day~0 onboarding, two free-living screening meals, daily indexed meals during the deployment, and the pre- and post-study interview.}

\begin{table*}[htbp]
\centering
\caption{Participant demographics and study assignment. Single earbud was worn on the dominant chewing side. All participants met our fast-eater screening criteria (chews per swallow $<20$ and food consumed per minute $> 50$ g/min). The sequence order was counterbalanced across participants.}
\resizebox{0.95\textwidth}{!}{
\begin{tabular}{c c c c C{3.0cm} C{4.0cm} c c}
\toprule
\multirow{2}{*}{\textbf{Participant id}} &
\multirow{2}{*}{\textbf{Gender}} &
\multirow{2}{*}{\textbf{Age}} &
\multirow{2}{*}{\makecell{\textbf{Dominant} \\ \textbf{chewing side}}} &
\multicolumn{3}{c}{\textbf{Baseline eating pace}} &
\multirow{2}{*}{\makecell{\textbf{Condition sequence}}} \\
\cmidrule(lr){5-7}
& & & &
{chews per swallow} &
{food consumed per min} &
\add{eating-episode duration}
\\
\midrule\midrule
P1&F&22 &Left & 11.78 &66.05 g/min & \add{10 min 2s} & Control $\rightarrow$ Experiment\\
P2&M&20 & Left & 14.23 &  67.55 g/min & \add{9 min 14s} &Control $\rightarrow$ Experiment\\
P3&M&24 &Right & 13.68 & 59.35 g/min & \add{9 min 3s} & Control $\rightarrow$ Experiment\\
P4&F&24 & Right & 16.72 & 53.56 g/min & \add{8 min 21s} & Experiment $\rightarrow$ Control\\
P5&M&23 &Left & 17.77 & 68.29 g/min & \add{8 min 9s} &Control $\rightarrow$ Experiment\\
P6&M&22 &Right & 11.22 & 70.85 g/min & \add{7 min 3s} & Control $\rightarrow$ Experiment\\
P7&M&20 &Right & 15.41 & 114.93 g/min & \add{5 min 53s} & Control $\rightarrow$ Experiment\\
P8&F&22 &Right & 9.04 & 139.18 g/min & \add{6 min 17s}   & Control $\rightarrow$ Experiment\\
P9&F&29 &Both & 14.75 & 64.88 g/min & \add{7 min 30s} & Experiment $\rightarrow$ Control \\
P10&M&29 &Both & 15.09 & 79.59 g/min & \add{8 min 22s} & Experiment $\rightarrow$ Control\\
P11&F&23 &Left & 16.56 & 72.53 g/min & \add{8 min 31s} & Experiment $\rightarrow$ Control\\
P12&M&23 &Left & 14.23 & 83.29 g/min & \add{9 min 8s} & Experiment $\rightarrow$ Control\\
P13&M&24 &Left & 14.21 & 78.23 g/min & \add{8 min 12s} & Experiment $\rightarrow$ Control\\
P14&M&24 &Right & 17.26 & 100.60 g/min & \add{6 min 57s} & Experiment $\rightarrow$ Control\\
\bottomrule
\end{tabular}}
\label{tab:participants}
\end{table*}

\subsubsection{Study Design}

We conducted a 13-day within-subject field study to evaluate the performance of the JIT intervention and the short-term carryover of our closed-loop system in daily eating contexts. The study compared two conditions:
\begin{itemize}
    \item \textbf{Control condition:} Participants received a pre-meal reminder at the beginning of each eating episode, with no intervention or feedback delivered during or after the meal.
    \item \textbf{Experiment condition:} Based on the closed-loop system of Earinter, participants received the pre-meal reminder and the JIT audio interventions delivered through the earbud during eating. After each meal, participants also received a post-meal summary report as feedback.
\end{itemize}

The Control condition represents a commonly used approach based on pre-meal self-cueing or external reminders. In contrast, the Experiment condition incorporates our JIT in-meal intervention design, and adds a post-meal summary to complete a per-meal closed loop. This design allows us to evaluate the full effect of Earinter, highlighting the value of a continuous intervention pipeline that intervenes during the meal and reinforces behavior change through post-meal reflection.

\begin{figure}[htbp]
    \centering
    \includegraphics[width=\linewidth]{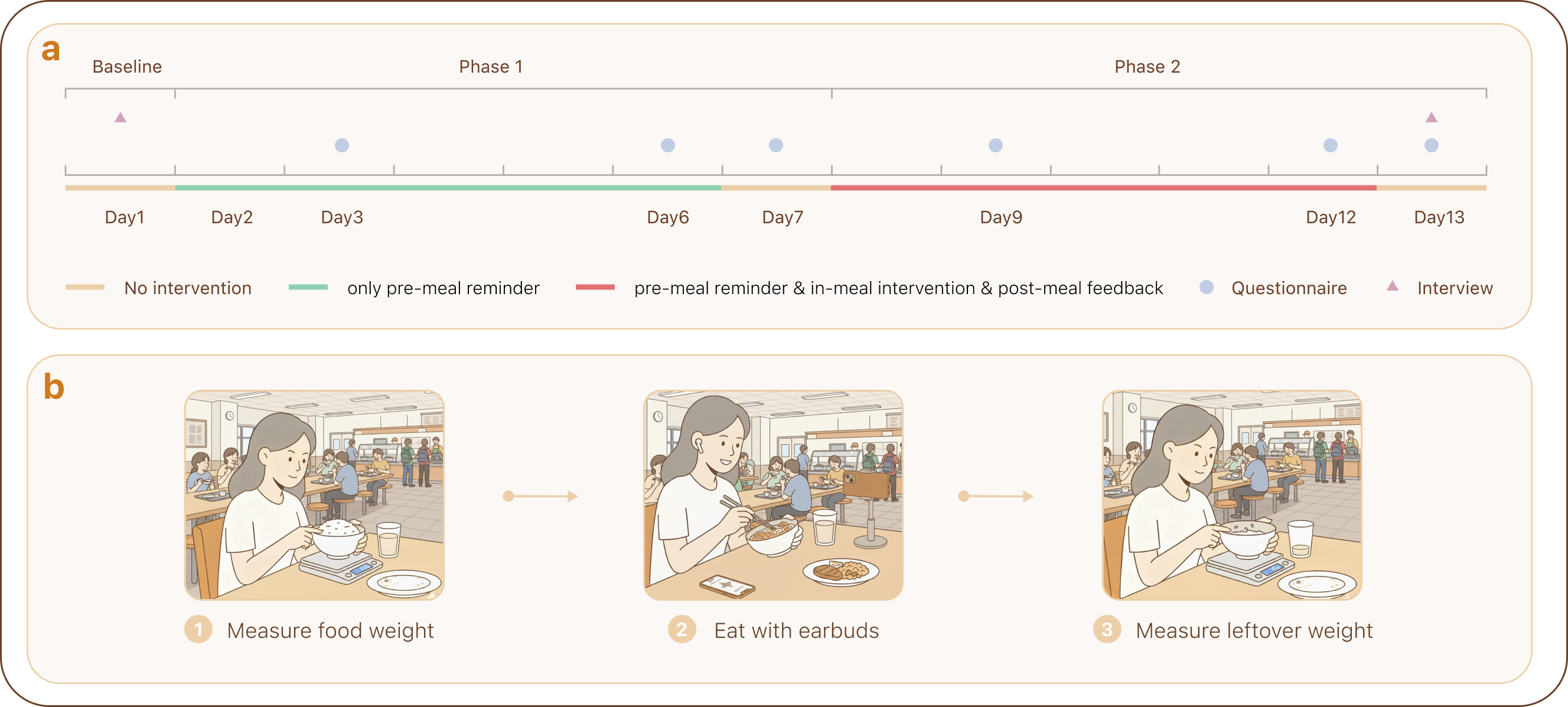}
    \caption{User study overview for \textit{\modify{Control} $\rightarrow$ Experiment}. (a) Timeline of the 13-day within-subject field study: a baseline day (no intervention), followed by two 6-day phases with five active days under each condition (Control: only pre-meal reminder; Experiment: pre-meal reminder \& in-meal JIT intervention \& post-meal feedback) and a retention day with no support. Dots and triangles mark questionnaire and interview. (b) Daily procedure for data collecting: participants \delete{measured body weight and fat, }measured food weight before eating, ate with earbuds, and weighed leftovers to estimate intake.}
    \label{fig:user_study}
\end{figure}

Beyond the condition design, we explicitly incorporated retention days at the end of each condition. This allowed us to quantify short-term behavioral persistence after withdrawing all support and reduce potential carryover contamination in our within-subject cross-over design. As shown in Fig.~\ref{fig:user_study}a, the overall study followed a three-stage pipeline:

\begin{itemize}
    \item \textbf{Baseline (Day 1):} Participants ate normally with the sensing and reasoning part of Earinter enabled, but received no audio prompts and feedback throughout the eating episode, including before and after eating.
    \item \textbf{Phase 1 (Days 2--7):} Five consecutive days under the assigned condition (Days 2--6), followed by a retention day (Day 7) with no prompts or feedback.
    \item \textbf{Phase 2 (Days 8--13):} Five consecutive days under the other condition (Days 8--12), followed by a retention day (Day 13) with no prompts or feedback.
\end{itemize}

To mitigate potential order effects, participants were randomly assigned to one of two sequences (\textit{Control $\rightarrow$ Experiment} or \textit{Experiment $\rightarrow$ Control}), resulting in a counterbalanced design across participants. Moreover, on Day~0, participants provided informed consent and received instructions on wearing the earbuds and using the Earinter app. To ensure consistent sensing and intervention delivery, they identified their dominant chewing side via self-report or a brief chewing calibration task and wore a single earbud on that side throughout the study. We further selected earbud tips to ensure a tight fit with each participant's ear canal.

The study protocol was approved by the institutional review board. Demographic and study design for each participant are summarized in Table \ref{tab:participants}.

\subsection{Data Collection and Preprocess}

\subsubsection{Measures and Data Collection}
\label{sec:measures}
During the whole study, we collected both objective behavioral measures and subjective feedback to evaluate system usability, intervention effects, and user experience in daily eating contexts.

\modify{Participants were asked to measure pre-meal food weights.} During meals, to support validation, participants recorded a rear-facing video. We also collected bone-conducted audio signals, including chewing and swallowing, via acoustic vibrations from the earbud. Then, we calculated food intake (excluding water) using post-meal food weights after finishing eating. These meal-level data were used to derive eating-pace indicators and to characterize how eating patterns evolved across different conditions. 
Daily procedures \modify{are} shown in Fig. \ref{fig:user_study}b.

To capture subjective eating experience and system usability over time, participants completed questionnaires on Days~3,~6,~7,~9,~12, and~13. The questionnaire package included (1) the \textit{Self-Reported Eating Experience Questionnaire} (Appendix \ref{appendix:daily_questionnaire}) and (2) the \textit{System Usability Scale (SUS)} \cite{brooke1996sus}. The eating-experience questionnaire assessed perceived changes in eating pace (e.g., chewing more \cite{fukuda2013chewing, sonoki2013effects}, slower eating), intake-related behaviors (e.g., reduced intake \cite{smit2011does}, smaller bites \cite{ruijschop2011effects, zijlstra2009effect, spiegel2000rate}) and overall meal experience.

Moreover, we conducted semi-structured interviews at the beginning of the study (Day~1) and after study completion (Day~13). The Day~1 interview focused on understanding how fast-eating behaviors are formed and maintained through the interaction of habit and context, and eliciting participants' perceived feasibility of intervention strategies under real-world constraints. The Day~13 interview focused on participants' eating-pace related behavioral changes across conditions, their subjective experience of the system, and recommendations for improving long-term use. The complete interview questions are shown in Appendix \ref{appendix:interview}.

\subsubsection{\modify{Data Preprocessing, Annotation, and Qualitative Coding}}
\label{sec:data_analysis}

All 14 \modify{remaining} participants completed the 13-day protocol and we collected 364 meal sessions. Prior to pre-processing and analysis of daily recordings and interviews, we screened the dataset based on participant self-reports and author inspection. \add{For food-consumption speed, the analyzed eating-episode duration was defined to reflect active ingestion time rather than the entire raw recording span. Pre-meal and post-meal non-eating segments (e.g., chatting before starting or after finishing the meal) were excluded. In the retained sessions, substantial non-ingestive interruptions during the meal were uncommon. When an individual session involved unusual circumstances, we reviewed the case and asked participants about the specific situation of that day during the post-study interview. We then determined whether the session should be treated as an anomalous meal and excluded, or whether a clearly non-ingestive segment should be removed during preprocessing. This procedure was intended to avoid inflating the denominator of food-consumption speed with time that did not reflect active ingestion.} After excluding 11 sessions (3\%) due to device issues (e.g., Bluetooth disconnections, corrupted audio files, or incomplete logs) and 11 sessions (3\%) for participant-reported anomalous meals (e.g., unusually time-constrained meals or other individual reasons), 342 valid meal sessions remained. \add{Across 14 participants over the 13-day study, this corresponds to an average of 24.43 valid sessions per participant, or approximately 1.88 valid sessions per participant per day.} Our conclusions were not materially affected by these exclusions.

We used frontal video recordings as the primary ground-truth source for ingestion-related events \cite{bedri2017earbit, bi2018auracle}. To derive meal-level chewing and swallowing counts, we recruited crowd annotators and performed manual annotation using ELAN \cite{ELAN2025}. Each video was independently annotated by two annotators. We computed Cohen's Kappa $\kappa$ for each video and required $\kappa \geq 0.75$ to indicate substantial agreement \cite{landis1977measurement,mordal2010norwegian}. For videos that did not meet this threshold, the first author reviewed the annotations, adjudicated disagreements, and finalized the labels in consultation with the research team. The finalized annotations were then aggregated to compute per-meal counts and eating-pace-related measures used in subsequent analyses.

We analyzed the semi-structured interviews using a bottom-up thematic analysis approach \cite{braun2006using}. Given the relatively small sample size, the first author conducted the full coding process, iteratively developing and refining codes and candidate themes. The broader research team subsequently reviewed the coding structure and the resulting themes. Then, we refined the final theme set through group discussion until consensus was reached. We did not compute inter-rater reliability. Consistent with qualitative research guidance, we prioritized collaborative interpretation and reflexive discussion over purely statistical agreement measures \cite{o2020intercoder}.

\subsubsection{\add{Statistical Analysis}}
\label{sec:statistical_analysis}

\add{The statistical analyses were organized around two primary analysis questions derived from Earinter's intended intervention mechanism. First, does the active \textit{Experiment} condition increase oral processing intensity, operationalized as chews per swallow (CPS), relative to the active \textit{Control} condition? Second, does the active \textit{Experiment} condition reduce food-consumption speed (g/min) relative to the active \textit{Control} condition? Accordingly, the two primary behavioral outcomes were CPS and food-consumption speed, and the primary comparison for both outcomes was the within-participant contrast between active \textit{Experiment} days and active \textit{Control} days. Comparisons involving \textit{Baseline} and retention-probe days were treated as secondary analyses intended to contextualize pre-intervention behavior and examine short-term persistence after support withdrawal. Subjective questionnaire factors were treated as secondary outcomes.}

\add{To make the within-subject nature of the design more transparent, we computed participant-level paired summaries comparing active Control and active Experiment days. For each participant, condition-level CPS was calculated as total chews divided by total swallows across valid meals within that condition, and condition-level food-consumption speed was calculated as total food consumed divided by total analyzed eating-episode duration across valid meals within that condition. We visualize these within-participant contrasts using paired plots and assess them using two-sided Wilcoxon signed-rank tests as robustness analyses.}

\add{Because the study followed a within-subject cross-over design, we used meal-level mixed-effects models that retain all valid sessions while accounting for repeated measures within participants, within-day clustering, and sequence effects. For CPS, we fit a Negative Binomial Generalized Linear Mixed Model (NB-GLMM) with chew count as the outcome and $\log(\mathrm{swallow\ count})$ as an offset, so that the model estimates chewing rate per swallow. Fixed effects included Condition (Baseline, Control, Control\_retention, Experiment, Experiment\_retention), Phase, and Condition sequence (Control$\rightarrow$Experiment vs.\ Experiment$\rightarrow$Control), with random intercepts for participants and participant-by-day random intercepts.}

\add{For food-consumption speed (g/min), our primary linear mixed-effects model (LMM) was fit on post-baseline meals to isolate differences among the study conditions after Day~1. Fixed effects included Condition (Control, Control\_retention, Experiment, Experiment\_retention), Meal type (lunch vs.\ dinner), Phase, and Condition sequence, with random intercepts for participants and participant-by-day random intercepts. Baseline food-consumption speed (Day~1) was included as a covariate. We report Type-III tests for omnibus effects and conducted post-hoc contrasts using estimated marginal means (EMMs) with Holm adjustment for multiple comparisons. We additionally report a complementary model that treats Baseline as an additional Condition level as a robustness check.}

\add{For self-reported eating-experience questionnaires, items were grouped into four theoretically derived factors: Behavioral Positive, Appetite, Fullness, and Comfort (see Appendix~\ref{appendix:daily_questionnaire}). Behavioral Positive was computed as a composite factor, whereas Appetite, Fullness, and Comfort were analyzed as ordinal item-level factors. We assessed internal consistency using Cronbach's $\alpha$ and report reliability values in the Results. Behavioral Positive was analyzed using an LMM, whereas Appetite, Fullness, and Comfort were analyzed using cumulative link mixed models (CLMMs) to respect the ordinal response scale. These models included fixed effects of Condition (Control vs.\ Experiment), Phase (Active vs.\ Retention), and their interaction, with random intercepts for participants.}

\add{
Because the active \textit{Experiment}-versus-\textit{Control} contrast was the primary comparison for the two behavioral outcomes, it was interpreted as the primary analysis contrast. Additional pairwise contrasts among condition states (e.g., involving \textit{Baseline} or retention-probe days) were treated as secondary, estimated using EMMs, and adjusted using Holm correction. Pairwise contrasts for questionnaire outcomes were estimated using EMMs and adjusted using Tukey correction. 
}

\subsection{Results}
\label{results}

\begin{figure}[h]
    \centering
    \includegraphics[width=.95\linewidth]{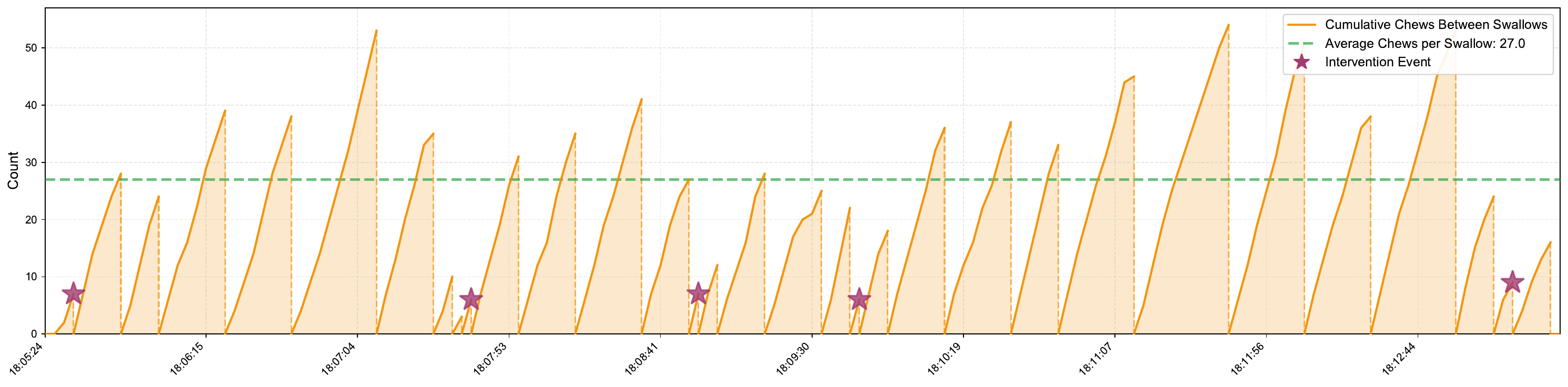}
    \includegraphics[width=.95\linewidth]{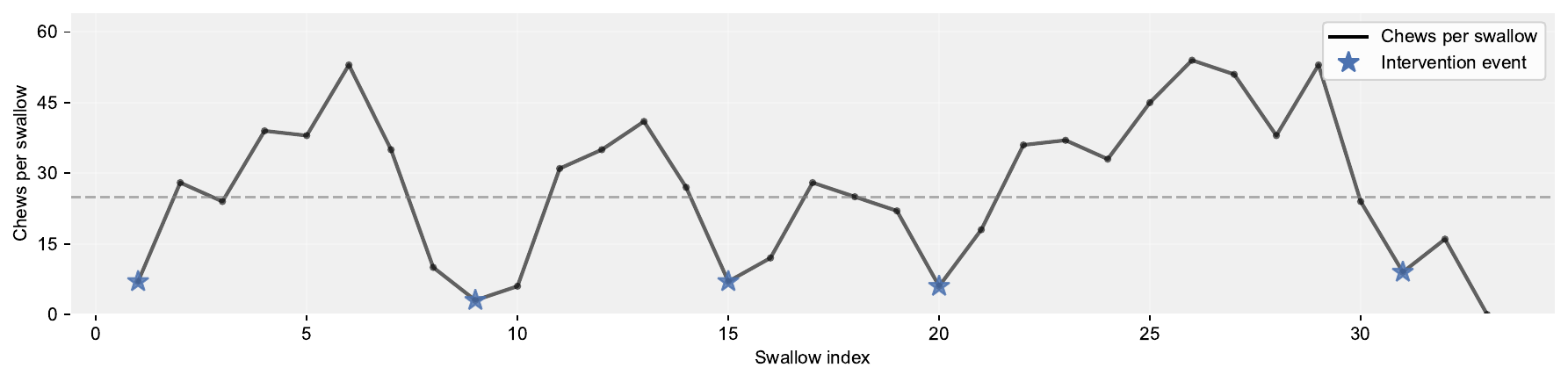}
    \caption{Representative eating-pace signals for P5 from the second day in \textit{Experiment} condition. The top plot shows time-based cumulative chewing activity, and the bottom shows chews per swallow. Star markers indicate moments when Earinter triggered JIT prompts based on the real-time estimated pace.}
    \label{fig:user_study_system_overview}
\end{figure}

To provide an intuition for how the Earinter system operates in situ, we present one representative day from the field study (P5, Day~2 in the \textit{Experiment} condition) in Fig.~\ref{fig:user_study_system_overview}. The corresponding eating-pace signals for all participants on that day are reported in Appendix~\ref{appendix:eating_pace}.

\subsubsection{Quantitative Results}

\paragraph{Participant-level paired comparison of active conditions}
\label{chap:paired_comparison}

\begin{figure}[h]
    \centering
    \includegraphics[width=0.9\linewidth]{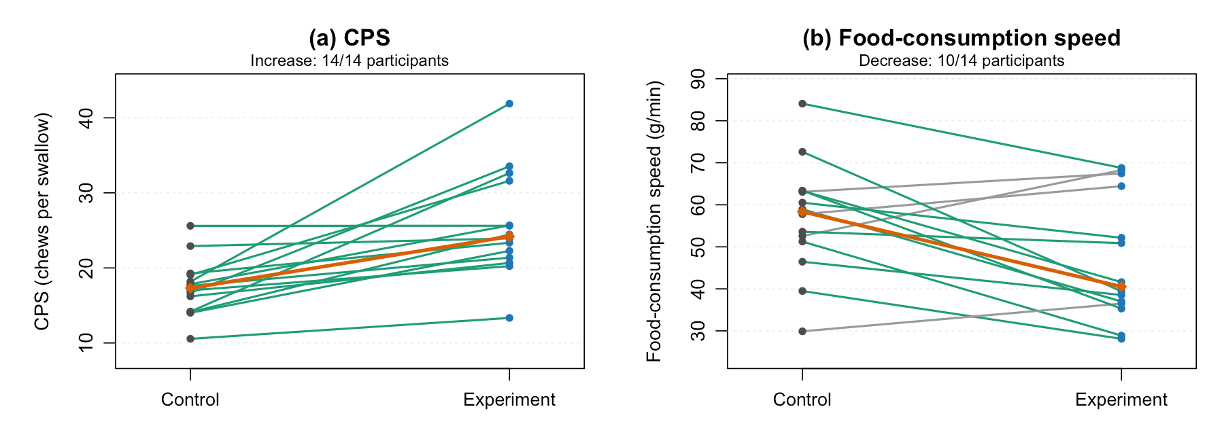}
    \caption{\add{Participant-level paired comparisons between active \textit{Control} and active \textit{Experiment} days. Each line represents one participant, with condition-level values aggregated across all valid active meals within that condition. (a) CPS, computed as total chews divided by total swallows. (b) Food-consumption speed, computed as total food consumed divided by total analyzed eating-episode duration. Green lines indicate changes in the expected direction (higher CPS or lower food-consumption speed under \textit{Experiment}), gray lines indicate changes in the opposite direction, and the orange line indicates the median across participants within each condition.}}
    \label{fig:paired_plot}
\end{figure}

\add{To make the within-subject cross-over design explicit, we first summarized each participant's active \textit{Control} and active \textit{Experiment} data. As shown in Fig.~\ref{fig:paired_plot}, CPS increased under \textit{Experiment} for all participants. At the participant level, CPS increased from a median of 17.31 under \textit{Control} to 24.20 under \textit{Experiment}, with a median paired difference of 6.15 CPS (Wilcoxon signed-rank test: $V=105$, $p=0.000122$).}
\add{Food-consumption speed decreased under \textit{Experiment} for 10/14 participants. The participant-level median decreased from 58.36 g/min under \textit{Control} to 40.52 g/min under \textit{Experiment}, with a median paired difference of -9.87 g/min (Wilcoxon signed-rank test: $V=18$, $p=0.0295$).}
\add{Although food-consumption speed showed more individual heterogeneity than CPS, all participants increased CPS under \textit{Experiment}. This divergence is plausible in free-living meals because food-consumption speed depends not only on oral processing intensity, but also on meal content, bite size, and the amount of food consumed per unit time. We therefore interpret CPS and food-consumption speed as complementary indicators of eating-pace regulation.} 
\add{The paired summaries above aggregate each participant's active \textit{Control} and active \textit{Experiment} meals into one value per condition, whereas the following descriptives and mixed-effects models use meal-level observations.}

\begin{figure}[htbp]
    \centering
    \includegraphics[width=\linewidth]{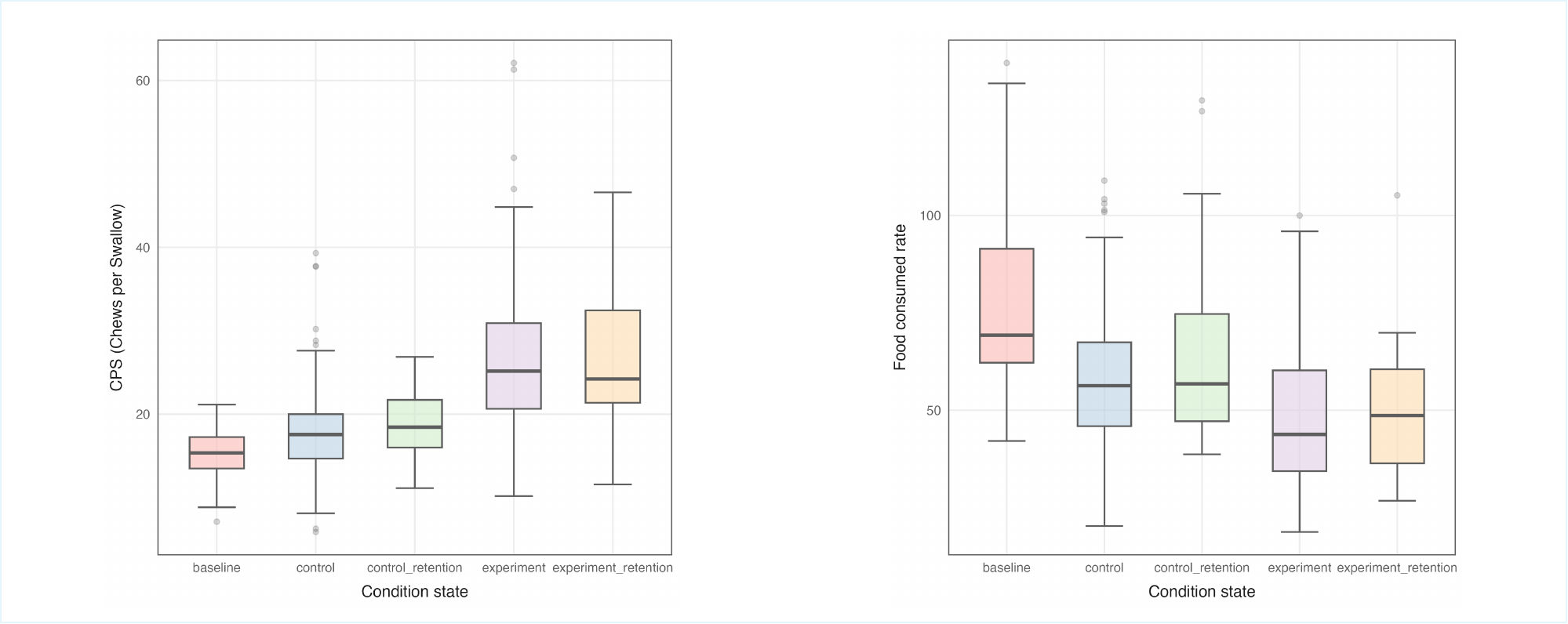}
    \caption{Box plots showing the distribution of \add{meal-level} eating pace metrics across five conditions: \textit{Baseline, Control, Control\_retention, Experiment, and Experiment\_retention}. The left plot shows the chews per swallow (CPS), indicating the oral process intensity. The right plot illustrates the food consumed rate (g/min), representing the speed of food intake.}
    \label{fig:eating_pace}
\end{figure}

\paragraph{Eating pace -- chews per swallow (CPS)}
\add{We analyzed meal-level CPS using the NB-GLMM described in Section~\ref{sec:statistical_analysis}, which retains all valid sessions while accounting for repeated measures, phase, and condition sequence.}
\modify{Meal-level descriptives, shown in Fig.~\ref{fig:eating_pace}, indicated that participants chewed on average 15.02 times per swallow at \textit{Baseline} ($SD = 3.67$). Under the active \textit{Control} condition, mean CPS was 17.88 ($SD = 5.32$), whereas under the active \textit{Experiment} condition, mean CPS increased to 26.42 ($SD = 8.61$). During retention-probe days, CPS was 18.60 ($SD = 4.20$) after \textit{Control} and 26.29 ($SD = 8.94$) after \textit{Experiment}.}

\modify{The primary analysis contrast showed that the active \textit{Experiment} condition produced significantly higher CPS than the active \textit{Control} condition (estimated effect = 0.37, $p < 0.0001^{****}$), indicating that closed-loop JIT prompting increased oral processing intensity relative to reminder-only support. In secondary retention-related contrasts, \textit{Experiment\_retention} remained higher than \textit{Control} (estimated effect = 0.32, $p < 0.001^{***}$). Relative to \textit{Baseline}, CPS increased significantly under \textit{Experiment} both during active intervention (estimated effect = 0.53, $p < 0.001^{***}$) and on the retention-probe day (estimated effect = 0.53, $p < 0.001^{***}$), whereas \textit{Control} and \textit{Control\_retention} showed only modest increases ($p < 0.05^{*}$). Finally, CPS did not significantly differ between \textit{Experiment} and \textit{Experiment\_retention} (estimated effect = 0.0023, $p \approx 1$), suggesting that the increase in oral processing intensity was maintained after support withdrawal within the study window. Together, these results indicate that Earinter more strongly regulated eating pace than reminder-only support, with short-term persistence on no-support probe days.}

\paragraph{Eating pace -- food consumed per minute}
\add{We analyzed meal-level food-consumption speed using the LMM described in Section~\ref{sec:statistical_analysis}.}
\modify{The primary model showed a significant main effect of \textit{Condition} on food-consumption speed 
($F(3,155.36)=15.11$, $p < 0.001^{***}$). For the primary analysis contrast, the active \textit{Experiment} condition showed a lower estimated food-consumption speed 
($EMM = 47.82$, $95\%CI = [41.39, 54.25]$) than the active \textit{Control} condition 
($EMM = 58.31$, $95\%CI = [51.89, 64.73]$; $\Delta = -10.48$ g/min, $t=-5.31$, $p < 0.001^{***}$). 
This result indicates that closed-loop JIT prompting reduced food-consumption speed relative to reminder-only support. In secondary retention-related contrasts, \textit{Experiment\_retention} remained lower than \textit{Control} 
($EMM = 49.01$, $95\%CI = [40.69, 57.33]$; $\Delta = -9.29$ g/min, $t=-2.71$, $p=0.02^{*}$). 
The difference between \textit{Experiment} and \textit{Experiment\_retention} was not significant 
($\Delta=-1.19$ g/min, $t=0.34$, $p=0.73^{ns}$), suggesting that the reduction in food-consumption speed was maintained on no-support retention-probe days within the study window.}

\modify{As a robustness check, we also fit a complementary LMM that treated \textit{Baseline} as an additional \textit{Condition} level. In this model, \textit{Baseline} food-consumption speed ($EMM = 76.40$, $95\%CI = [67.47, 85.33]$) was significantly higher than all other conditions, including \textit{Control} ($\Delta=18.05$ g/min, $p<0.0001^{****}$), \textit{Control\_retention} ($\Delta=11.32$ g/min, $p=0.0489^{*}$), \textit{Experiment} ($\Delta=28.55$ g/min, $p<0.0001^{****}$), and \textit{Experiment\_retention} ($\Delta=27.30$ g/min, $p<0.0001^{****}$). Together, these results support the conclusion that Earinter reduced food-consumption speed beyond reminder-only support, with short-term persistence after support withdrawal.}

\paragraph{Self-reported eating experience}

\modify{We next report the self-reported eating-experience factors analyzed using the models described in Section~\ref{sec:statistical_analysis}. The questionnaire items were grouped into four factors: \textit{Behavioral Positive} (positive perception of eating behavior), \textit{Appetite}, \textit{Fullness}, and \textit{Comfort} (wearing comfort). A total of 84 valid observations were included (28 \textit{Control}, 28 \textit{Experiment}, 14 \textit{Control\_retention}, and 14 \textit{Experiment\_retention}). The overall scale showed good internal consistency (Cronbach's $\alpha = 0.866$), and the \textit{Behavioral Positive} subscale showed excellent consistency ($\alpha = 0.929$). Results are summarized in Fig.~\ref{fig:interv_result}; see Appendix~\ref{appendix:daily_questionnaire} for the mapping of questionnaire items to factors.}



\begin{figure}[htbp]
  \centering
  \includegraphics[width=1.0\linewidth]{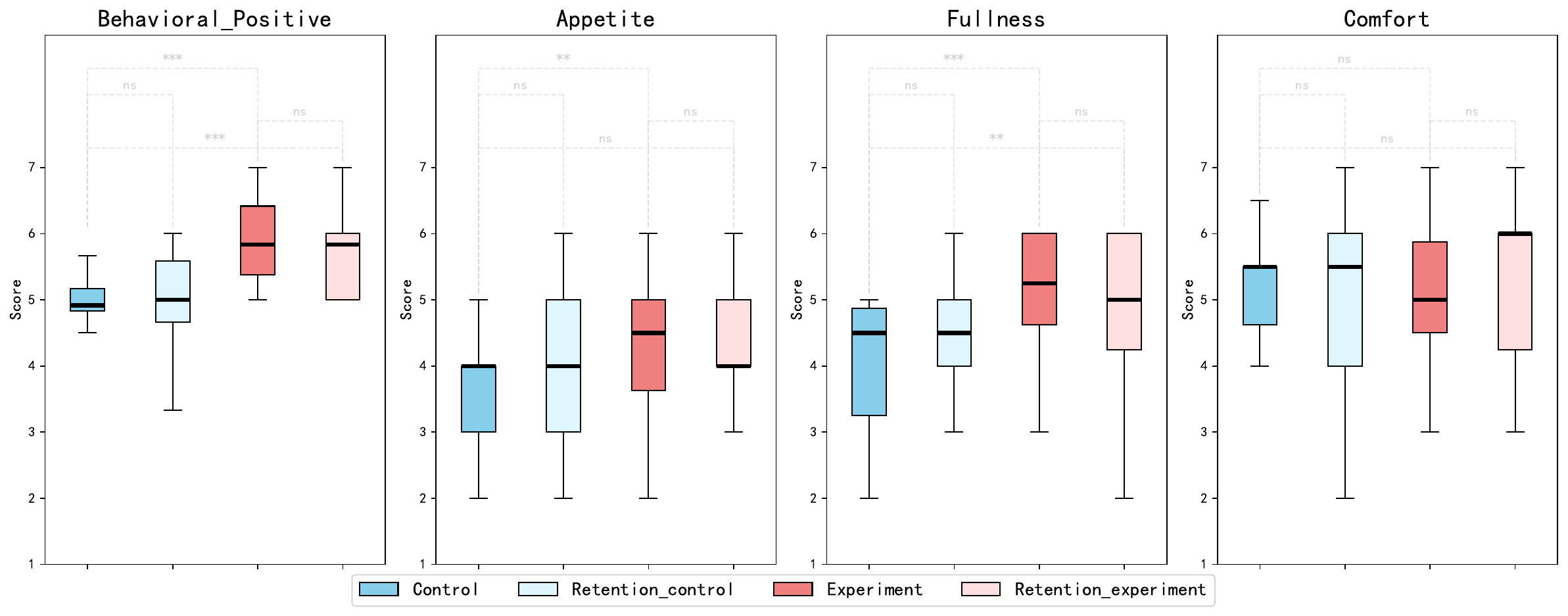}
  \caption{Box plots of self-reported eating experience scores for the four factors, \textit{Behavioral Positive}, \textit{Appetite}, \textit{Fullness}, and \textit{Comfort}, across the \textit{Control, Control\_retention, Experiment, and Experiment\_retention} conditions. Asterisks denote statistically significant pairwise differences (* indicates $p$ < 0.05, ** indicates $p$ < 0.01, *** indicates $p$ < 0.001).}
  \label{fig:interv_result}
\end{figure}

{Behavioral Positive:}
\modify{Participants perceived their eating behavior as more positive during the \textit{Experiment} condition ($M$ = 5.68, $SD$ = 1.12) than the \textit{Control} condition ($M$ = 4.88, $SD$ = 0.83).}
An LMM analysis confirmed this difference ($\beta = 0.80$, $SE = 0.13$, $t = 6.12$, $p < .001^{***}$), indicating improved perceived eating control under the JIT intervention. \modify{After support withdrawal}, \textit{Experiment\_retention} session still yielded higher scores ($M$ = 5.55, $SD$ = 1.03) than \textit{Control} ($p = 0.0001^{***}$), \modify{consistent with short-term persistence of perceived behavioral improvement}, although scores were negligibly lower than during the Intervention phase ($p = 0.415^{ns}$). These results indicate the short-term persistence of perceived behavioral improvement \modify{, with only mild attenuation after support withdrawal.}

{Appetite:}
Participants reported higher appetite-related scores during the \textit{Experiment} condition ($M$ = 4.32, $SD$ = 1.25) than during \textit{Control} ($M$ = 3.68, $SD$ = 1.09).
\modify{The difference was confirmed with a CLMM analysis ($OR = 4.43$, $95\% CI = [1.54, 12.77]$, $p = 0.0059^{**}$), indicating improved perceived appetite regulation during the JIT intervention. No significant difference was observed between the \textit{Experiment} and \textit{Experiment\_retention} conditions ($p = 0.593^{ns}$), suggesting that this perceived improvement did not substantially decline after support withdrawal. Scores in \textit{Experiment\_retention} were slightly higher than those in \textit{Control}, although this comparison did not reach conventional significance 
($p = 0.056$). Together,  these results suggest that appetite-related perception improved during the JIT intervention in \textit{Experiment} condition and showed signs of short-term persistence after support withdrawal.}

{Fullness:}
Compared with \textit{Control} ($M = 4.07$, $SD = 1.12$), participants reported greater fullness during the \textit{Experiment} condition ($M = 5.00$, $SD = 1.25$). \modify{The CLMM showed that this difference was significant ($OR = 12.62$, $95\% CI = [3.76, 42.32]$, $p < .001^{***}$). After support withdrawal, \textit{Experiment\_retention} did not differ significantly from the active \textit{Experiment} condition ($p = 0.684^{ns}$), but remained significantly higher than \textit{Control} ($p = 0.0012^{**}$). These results suggest short-term persistence of perceived fullness improvement following the JIT intervention.}

{Comfort:}
The comfort factor was reverse-coded prior to analysis. Ratings were comparable between \textit{Control} ($M$ = 5.10, $SD$ = 1.07) and \textit{Experiment} ($M$ = 4.96, $SD$ = 1.37). \modify{No} significant condition effect was observed ($OR = 0.92$, $95\% CI = [0.34, 2.51]$, $p = 0.873^{ns}$). 
After \add{support} withdrawal, both comparisons remained non-significant (\textit{Control\_retention}: $p = 0.606^{ns}$; \textit{Experiment\_retention}: $p = 0.565^{ns}$). 
These results indicate that the presence of auditory feedback, whether JIT or not, did not reduce comfort relative to simply wearing the earbuds. Descriptively, \modify{the overall comfort rating ($M = 5.03$, $SD = 1.27$) suggests generally favorable comfort of commodity earbuds across phases.}

\paragraph{System Usability Scale.}

We evaluated the system's perceived usability using the System Usability Scale (SUS) \cite{brooke1996sus}. Given the documented advantage of 7-point Likert scales in enhancing response sensitivity \cite{finstad2010response}, we adopted this format for the SUS questionnaire following \citet{lee2024fluid}. 
Across \textit{Experiment} phases, the mean SUS score was 70.0 $\pm$ 9.6 ($median = 70.0$, $IQR = 16.7$),
placing the system within the "OK-Good" range on the interpretive scale proposed by \citet{bangor2008empirical}. 

Across the ten SUS items of experiment, participants reported higher ratings on positively worded statements such as "I found the system easy to use" ($M = 5.58$) and "I felt very confident using the system" ($M = 5.69$). These indicated improved perceived ease of use and confidence with the system. In contrast, lower scores appeared on negatively worded items such as "I found inconsistencies in the system" ($M = 2.84$) and "I needed to learn a lot of things before I could get going with this system" ($M = 2.78$), suggesting that users in the experimental condition perceived fewer complexities and a smoother learning process. These differences reflect a generally more intuitive and user-friendly experience in the experimental interface. 

Overall, participants perceived the earbuds-based JIT intervention system as usable and well-designed, indicating promising usability and user acceptance for real-world behavioral intervention contexts.

\subsubsection{Qualitative Findings}



We report qualitative findings from two rounds of semi-structured interviews, analyzed using an inductive thematic analysis approach (Section~\ref{sec:data_analysis}). Baseline interviews (Day~1) characterized participants' fast-eating routines and constraints that limit intentional change, while exit interviews (Day~13) captured experiences across conditions, perceived behavior change (including retention-probe days), and design suggestions. We synthesize themes across both rounds to explain why fast eating persists and how participants experienced our interventions in daily ingestion.

\paragraph{Theme 1: Habit- and context-driven fast eating makes self-regulation difficult, motivating low-burden external support.}
In baseline interviews, participants commonly described fast eating pace as an ingrained and automatic habit that developed and reinforced by some constraints over time. For instance, P12 noted that he had "always chewed less and eaten faster since childhood". Others attributed their pace to long-term time scarcity and efficiency-seeking (e.g., P5:"I want to save time."), including routines shaped by short breaks earlier in life.
Beyond habit, participants emphasized that pace was strongly context-dependent, varying with workload, social settings, and concurrent activities. P7 reported speeding up to match peers when eating together, while P14 described time-compressed lunch breaks that made slow eating feel unrealistic. Although some participants occasionally recognized fast eating pace, they reported that self-regulation was difficult to sustain. Because maintaining self-reminders felt cognitively demanding (e.g., P3) and sometimes physically uncomfortable during busy or stressful meals (e.g., P1, P8). These barriers motivated a preference for external and low-burden support, so participants expressed interest in prompts triggered by objective eating signals. They also cautioned that messages which are overly long or delivered too frequently could increase perceived burden and become easier to ignore (e.g., P3, P6, P8), underscoring the importance of pilot study in intervention design (Section \ref{chap:pilot}).

\paragraph{Theme 2: Participants attributed fast eating to delayed satiety. Slowing down increased perceived fullness, while intake was constrained by portion norms.}
Participants described a perceived mismatch between fast eating pace and internal satiety awareness. Baseline interview suggested heterogeneity in how pace relates to intake. Some participants described relatively fixed portions, whereas others (e.g., P3, P8, P11) linked fast eating to delayed satiety cues and a tendency to overeat. Several participants described this experience as "satiety-related signals cannot keep up when eating too fast".
In exit interviews, multiple participants reported that when the closed-loop condition helped them chew more and slow down, they felt full earlier or found it harder to finish their usual portion (e.g., P3, P8). However, regardless of increased fullness, many participants reported little change in total intake. They often cite plate-clearing norms (avoiding food waste) and fixed meal expectations as constraints that decouple pace from intake. Taken together, these results suggest that pace interventions may more reliably change chewing dynamics and perceived satiety than absolute intake for users whose portions are socially or habitually fixed.

\begin{table*}[t]
\centering
\caption{Summary of participants' self-perceived behavioral changes and design preferences reported in semi-structured interviews (N=14). Changes were aggregated per participant ($\checkmark$ = explicitly mentioned; -- = not mentioned). "Condition preference" indicates overall preference among baseline, control and experiment (closed-loop) conditions; "Preference for JIT prompt type" refers to the prompt framing participants favored (see Section \ref{chap:jit_content}).}
\resizebox{\textwidth}{!}{
\begin{tabular}{cccccccccc}
\toprule
\multirow{2}{*}{\textbf{Participant id}} &
\multicolumn{6}{c}{\textbf{Self-perceived behavioral changes}} &
\multicolumn{2}{c}{\textbf{Design preferences and attitudes}}
\\
\cmidrule(lr){2-9}
 & Slower eating pace & Increased CPS & Smaller bite size & Earlier fullness  & Reduced food intake & Heightened food awareness & Condition preference & JIT prompt preference \\
\midrule\midrule
P1 & $\checkmark$ & $\checkmark$ & -- & -- & -- & -- &Experiment & Control Theory \\
P2 & $\checkmark$ & $\checkmark$ & -- & -- & -- & -- &Experiment & Control Theory \\
P3 & $\checkmark$ & $\checkmark$ & -- & $\checkmark$ & $\checkmark$ &$\checkmark$ &Experiment &  Goal Framing Theory\\
P4 & $\checkmark$ &$\checkmark$ & -- & -- & -- &$\checkmark$ &Experiment &  Control Theory\\
P5 & $\checkmark$ & $\checkmark$ & $\checkmark$ & -- & -- & $\checkmark$ &Experiment &  Control Theory\\
P6 & $\checkmark$ & $\checkmark$ & -- & -- & -- & $\checkmark$ &Experiment &  Goal Framing Theory\\
P7 & $\checkmark$ & $\checkmark$ & -- & -- & -- & $\checkmark$ &Experiment & Control Theory \\
P8 & $\checkmark$ & $\checkmark$ & -- & $\checkmark$ & $\checkmark$ & -- &Experiment &  Control Theory\\
P9 & -- & -- & -- & -- & -- & $\checkmark$ &Control &  Control Theory\\
P10 & $\checkmark$ & $\checkmark$ & $\checkmark$ & -- & -- & $\checkmark$ &Experiment &  Control Theory\\
P11 & $\checkmark$ & $\checkmark$ & $\checkmark$ & $\checkmark$ & -- & $\checkmark$ &Experiment &  Control Theory\\
P12 & $\checkmark$ & -- & $\checkmark$ & -- & -- & -- &Baseline & Dual System Theory (System~1)\\
P13 & $\checkmark$ & $\checkmark$ & -- & -- & -- & -- &Experiment & Control Theory \\
P14 & $\checkmark$ & $\checkmark$ & $\checkmark$ & $\checkmark$ & -- & $\checkmark$ &Experiment &  Goal Framing Theory\\

\bottomrule
\end{tabular}}
\label{tab:part_interview_result}
\end{table*}

\paragraph{Theme 3: JIT interventions were most effective when actionable and tightly coupled to eating pace, supporting moment-to-moment correction.}
Baseline interviews indicated that participants were generally optimistic about our closed-loop system including JIT intervention. For exit interview, as shown in Table~\ref{tab:part_interview_result}, most participants preferred closed-loop system and reported that the closed-loop condition effectively helped slow their moment-to-moment eating pace to a stable level. The most frequently mentioned behavioral change was increased chewing per swallow. For example, P1 explicitly stated that "the most noticeable positive effect was chewing more". 
Some participants also reported learning-related changes over time, such as "gradually adopting smaller bites later in the final period" (P10) and "eating in smaller and more even bites" (P5). Participants further described carryover in retention day and after the user study. For instance, P10 noted that even without continuing to use the system after the study, "this experience really made me eat more slowly". These accounts are consistent with our intended mechanism, in which brief in-the-moment interruptions redirect attention to slower eating pace. Together, they suggest early signs of behavioral persistence beyond active prompting, aligning with the long-term aspirations of closed-loop system design.

Participants also differentiated between prompt styles. Across interviews, many expressed a preference for prompts that provided actionable, control-oriented guidance, because such prompts supported immediate self-correction. In contrast, messages derived from more general behavioral theories were often described as generic, and goal-framing prompts were sometimes perceived as "too common-sense" (P11) or "not relevant to me" (P4). 

\paragraph{Theme 4: Participants desired positive reinforcement, transparency, and personalization to support long-term use.}
Participants suggested incorporating more positive reinforcement to enhance motivation and willingness to persist. For example, P2 and P5 recommended that the system acknowledge improvements when they returned to a healthier pace. In addition, about half of participants explicitly requested greater transparency and personalization. Transparency referred to helping users better understand their ongoing eating process through richer, real-time indicators (e.g., P4, P5: a progress-bar-like visualization) rather than a "black-box" experience. Personalization requests focused on adapting targets to meal content and individual differences. They noted that different food properties likely require different chewing expectations (e.g., P12), and suggested allowing users to set pre-meal intentions or goals (P1). Together, these suggestions point to design directions that could improve acceptability and support sustained behavior change beyond the study period.

\section{DISCUSSION}

\subsection{Closed-Loop System for Dynamic Ingestive Behavior Regulation}

During a meal, eating pace can fluctuate dynamically without the eater's awareness, making purely open-loop strategies inherently misaligned with moment-to-moment changes~\cite{faltaous2021wisdom, hori2023masticatory}. Feedback may arrive too early or too late, or without sufficient context to be actionable, which can weaken perceived usefulness and lead users to disengage. We therefore adopt a closed-loop design that tightly couples sensing, real-time reasoning, and intervention~\cite{fang2025review}. In Earinter, continuous CPS estimation enables behavior-contingent support: the system intervenes only when inferred pace drifts toward fast eating and remains silent when pace is already within a desired range, reducing unnecessary interruptions while keeping feedback directly relevant to the user's immediate state. This design further enables an iterative \textit{$Observe \rightarrow Correct \rightarrow Observe$} cycle: Earinter cues correction at the moment of deviation, re-assesses the subsequent response through continued sensing, and re-prompts if the pace drifts again.

The closed-loop structure, as Earinter illustrates, maps naturally onto \textit{Prompt}, \textit{Ability}, and \textit{Motivation}~\cite{fogg2019fogg}. First, cues serve as timely \textit{Prompt}, triggered specifically at deviation moments rather than on a fixed schedule. Second, the closed-loop cycle strengthens perceived \textit{Ability} in two complementary ways. By externalizing sustained self-monitoring, the system lowers the cognitive effort required to notice and correct pace changes, making regulation easier in the moment. In addition, when users correct their behavior and the system subsequently remains silent, this immediate feedback serves as confirmation that their action led to measurable improvement, reinforcing a sense of self-efficacy and control. Third, the system supports \textit{Motivation} through a \modify{lightweight intervention policy designed} that reduces frustration and makes sustained regulation more likely, and through repeated, attainable corrections that help users experience progress. Although a two-week deployment cannot establish durable long-term habit change, we observed significant retention on no-support retention-probe days within the study window, suggesting early transfer from external prompting to self-regulation. As P10 reflected a week after the study, "I found the intervention really effective. I genuinely ate more slowly!" These findings suggest that closed-loop systems can iteratively reshape habitual eating pace through repeated, contextually grounded feedback. We encourage future HCI work on habit change to explicitly consider these iterative benefits and examine how repeated closed-loop systems may support cognitive-level internalization and persistence over longer time horizons.

\subsection{Earables Enable End-to-End Regulation as a Practical Platform}
\label{7_2}

\modify{Earables offer a practical substrate for closed-loop systems not only because they co-locate sensing and feedback, but also because ear-worn devices are already integrated into everyday life. Prior work characterizes earables as building on an existing consumer platform and notes the large-scale adoption of wireless earphones \cite{roddiger2022sensing}.}

\modify{Beyond convenience, earables provide several advantages as an intervention channel. In-ear audio cues are private and eyes-free, and therefore do not compete with visual attention during meals \cite{roddiger2022sensing, stanke2023can}. Compared with screen-based prompts, visible wearables, or tabletop systems, earables can deliver subtle guidance with relatively low social intrusiveness \cite{xu2020earbuddy}. Earinter further limits disruption through behavior-contingent triggering, cooldown intervals, and theory-grounded prompts, which together aim to preserve intervention salience while maintaining comfort and usability.}

\modify{The ear is also a favorable site for robust, fine-grained ingestive sensing \cite{shin2022mydj}. Compared with air-conducted microphones that are more susceptible to environmental noise in real meals, bone-conducted sensing captures chewing-related vibrations transmitted through the ear canal and skull, making the signal less sensitive to ambient sounds and more privacy-preserving. While prior ear-based systems often focus on coarse meal or eating-episode detection~\cite{bedri2017earbit, bi2018auracle, shin2022mydj}, Earinter targets within-meal pacing support and therefore requires chew-level reliability for real-time intervention.}

\modify{Building on these properties, Earinter repurposes a commodity earbud's bone-conduction voice sensor to capture chewing-related vibrations while using the same device to deliver in-ear prompts. This dual use enables a closed-loop system on an already-adopted form factor, reduces setup friction relative to multi-device configurations, and lowers the barrier to sustained daily use. Our evaluation provides further evidence for the practical value of this design. Comfort remains comparable to the control condition, with a mean rating of $5.03 \pm 1.27$, and the mean SUS score is $70.0 \pm 9.6$. Participants also report strong willingness to use the system frequently in the SUS ($M = 5.14, SD = 1.37$), suggesting that the earable form factor is generally comfortable and usable across diverse in-the-wild meals. Qualitative feedback from Wizard-of-Oz pilot sessions across solo, dyadic, and family co-present meals further informs the selected prompting strategy, favoring interventions that participants perceive as acceptable and minimally interruptive, with few reported negative reactions. Moreover, during post-study interviews, most participants indicate that the earbuds "do not impose noticeable burden during in-the-wild eating" (P2). Some participants routinely wear earbuds during meals (e.g., P6, P13) and report no perceived burden. In addition, several participants express a desire to continue using the system on their own earbuds (P9, P13), further supporting the practical advantages of earables.
Taken together, these findings suggest that this earable form factor can support low social and cognitive interference in many everyday meal contexts. }

\subsection{Towards the Just-In-Time Adaptive Intervention}
Just-in-time (JIT) shifts intervention from static advice on what to do to situated support about what to do now, and JIT adaptive intervention (JITAI) further emphasizes that effective support requires not only timely triggering but also adaptive decisions as context conditions change \cite{nahum2016just}. This framing is particularly suitable for habitual behaviors that fluctuate within daily episodes, such as eating pace, where a user's vulnerability may emerge intermittently and is hard to notice through sustained self-monitoring.  

We found that adaptivity is central to both effectiveness and acceptability. Earinter therefore treats the delivery policy as part of the intervention: pacing-contingent triggering, a cooldown interval, and varied prompt lengths together adjust the intervention dose to preserve relevance while reducing interruption. This is consistent with evidence that interventions embedded at in-situ decision points and requiring lightweight engagement can outperform static notifications in real-world behavior change~\cite{xu2022typeout}. In our study, participants also preferred actionable, control-theory-oriented prompts (e.g., "chew xx more times to reach the goal"), suggesting that JIT content should emphasize concrete, immediately executable actions. This aligns with a control-theoretic view of regulation --  users benefit from feedback that makes the current deviation legible and specifies how to correct it.

Looking forward, a promising direction is to extend earable-based pacing support toward fuller JITAI capabilities by incorporating richer contextual signals, such as social context, concurrent media use, time pressure, and personalization. These directions point toward JIT interventions that behave less like fixed reminders and more like adaptive companions for self-regulation of habitual behaviors.

\subsection{\add{Practical Reliability under Non-Ingestive Daily Activities}}
\label{sec:confound_stress_test}
\add{Beyond the meal-based evaluation, we also examined how the trained chewing-detection pipeline behaves under several non-ingestive activities during daily earbud use, with full task-level statistics reported in Appendix~\ref{app:non_ingestive_stress_test}. We recruited 10 participants to complete continuous no-eating blocks covering walking-related activities, speech and social activities, head motion, and audio playback. The results suggest that the detector is not indiscriminately activated by routine earbud use. Passive external speech, phone checking, and audio playback were rarely confused with chewing.
More specifically, within the speech/social block, self-generated speech introduced measurable false positives -- loud reading produced 3.851 FP/min, dyadic conversation produced 2.474 FP/min, and listening produced 0 FP/min.}

\add{At the same time, the additional test reveals clear boundary conditions. Continuous loud self-generated speech, repeated head motion, and stronger body movement such as jogging or fast stair walking produced non-negligible false positives, likely because bone-conduction sensing can also capture self-vocalization and motion-related vibrations. These results help define Earinter's practical operating envelope. 
Because Earinter is a user-initiated meal-time intervention rather than a passive all-day detector, users are expected to initiate it during typical meals rather than during high-motion activities or socially intensive meal contexts. Though, loud speech remains a realistic boundary condition. 
Future systems should therefore incorporate richer contextual suppression so that closed-loop support can remain effective and socially acceptable across a wider range of daily situations.}

\subsection{\modify{Limitations} and Future Work}
\label{7_4}




\paragraph{Deployment scope and participant diversity}

\modify{Our current study deployment lasted 13 days and primarily served to validate the feasibility and effectiveness of an in-the-wild closed-loop system for eating-pace regulation. Although we observed significant improvements in CPS and slower eating pace, together with short-term retention on no-support retention-probe days within the study window, this duration is still insufficient to establish durable habit change. Future work should conduct longer deployments to examine how effects evolve over time, including patterns of persistence, attenuation, and habituation. Importantly, these short-term effects were limited to eating-pace regulation and should not be interpreted as evidence of weight loss. In an exploratory participant-level analysis, the reduction in food-consumption speed was not significantly correlated with short-term body-weight change, irrespective of participants’ starting body weight. Longer studies with pre-specified intake and weight outcomes are needed to examine this relationship. Our participant sample is also limited in demographic diversity. Although participants included undergraduate students, graduate students, and campus workers, all enrolled participants were young adults aged 20--29. In addition, we did not conduct formal clinical screening for eating disorders using a validated assessment instrument, and instead relied on self-report during recruitment. Future work should evaluate the system with more diverse populations and incorporate more structured screening procedures when appropriate.}

\paragraph{Pace-estimation precision and meal-dependent targets.}

\modify{The current CPS estimation also has room for improvement. Because swallowing signatures are relatively weak in commodity earbud bone-conduction signals, Earinter infers swallow boundaries from inter-chew intervals rather than directly sensing each swallow. This limitation is reflected in the MAE. When prolonged inter-chew gaps reflect non-ingestive pauses rather than true swallows, the heuristic may insert false swallow boundaries and consequently underestimate CPS. In the analyzed field-study sessions, substantial non-ingestive interruptions were uncommon. The intervention policy also operates on continuously updated CPS estimates with pace thresholds and cooldown control, reducing the likelihood that isolated pause-related errors lead to persistent or excessive prompting. Advances in hardware or signal processing could enable more precise behavioral regulation while preserving a similarly portable and practical form factor. Healthy eating pace is also food-dependent. We used commonly cited chewing heuristics as an intuitive reference to standardize prompt design and evaluate system effects~\cite{alexander1998news, sonoki2013effects}. A promising direction for future work is to incorporate meal characteristics, such as food texture or type, to adapt CPS goals and improve robustness in daily use.}

\paragraph{Situational fit and social acceptability.}

\modify{Although our intervention policy is designed to limit interruption, it may still be misaligned with users' situational preferences in some contexts. In particular, in conversation-centered social meals, the acceptability of wearing earphones and receiving auditory prompts may depend on individual preferences, relationship norms, and the salience of the social setting. Future systems could therefore incorporate context-aware suppression for conversation-heavy moments, richer personalization of when and how prompts are delivered, or alternative device forms that are more robust to social sensitivity.}

\section{CONCLUSION}

In this paper, we present Earinter, a closed-loop system using commodity earbuds for regulating eating pace during daily meals. Earinter repurposes an earbud's bone-conduction voice sensor to detect chewing and infer swallow boundaries for real-time chews-per-swallow (CPS) tracking. With hours of data collected equally across sessions \modify{in quiet and noisy environments}, Earinter achieved reliable chewing detection ($F1 = 0.97$) and accurate pace estimation ($MAE = 0.18 \pm 0.13$ chews/min; 3.65 $\pm$ 3.86 chews/swallow). 
Guided by Dual Systems Theory, Earinter delivers theory-based just-in-time audio prompts with an adaptive delivery policy designed to \modify{preserve comfort and limit interruption}.
We conducted a 13-day within-subject field study with 14 participants to evaluate the effectiveness and acceptability of the closed-loop intervention in the wild. 
Our results show that Earinter significantly increases CPS (15.02 $\pm$ 3.67 $\rightarrow$ 26.42 $\pm$ 8.61) and slowed meal-level eating pace compared to self-cueing or external reminder, with significant retention on no-support retention-probe days and \modify{favorable comfort and usability.} However, these short-term pace-related changes were not evidence of body-weight loss.
Together, these results highlight the promise of commodity earables as practical platforms for theory-driven closed-loop JIT interventions that regulate eating pace in real-world settings.

\begin{acks}
This work is supported by the Natural Science Foundation of China under grant No. 62472244, 62132010, Key R\&D Program of Shandong Province of China under grant No. 2026CXGC010116, the Qinghai University Research Ability Enhancement Project under grant No. 2025KTSA05, Institute for Artificial Intelligence of Tsinghua University (THUAI), Undergraduate / Graduate Education Innovation Grants of Tsinghua University, and Tsinghua University Initiative Scientifc Research Program.
\end{acks}

\bibliographystyle{ACM-Reference-Format}
\bibliography{ref.bib}

\newpage
\appendix
\section{STATISTICAL DETAILS OF THE DATASET FOR EVALUATING REAL-TIME EATING PACE SENSING METHODS}

We summarized the eating acoustic events for each participant in both quiet (lab conference room) and noisy (cafeteria) scenarios, as shown in Table~\ref{tab:eating-dataset-appendix}, to evaluate our proposed real-time eating pace sensing method.

\begin{table}[H]
    \centering
    \small 
    \renewcommand{\arraystretch}{0.85}
    \caption{Per-participant Acoustic Events for Evaluating Real-time Eating Pace Sensing}
    \label{tab:eating-dataset-appendix}
    \begin{tabular}{llccc}
        \toprule
        \textbf{User ID} & \textbf{Environment} & \textbf{Duration (s)} & \textbf{Chewing Counts} & \textbf{Swallowing Counts} \\ \midrule
        
        \multirow{2}{*}{D1} & Quiet & 1296.53 & 2514 & 109 \\
                           & Noisy & 1410.59 & 2392 & 112 \\ \midrule
        \multirow{2}{*}{D2} & Quiet & 1620.08 & 1743 & 72 \\
                           & Noisy & 641.52 & 888 & 36 \\ \midrule
        \multirow{2}{*}{D3} & Quiet & 829.65 & 997 & 51 \\
                           & Noisy & 465.15 & 516 & 45 \\ \midrule
        \multirow{2}{*}{D4} & Quiet & 1128.75 & 879 & 48 \\
                           & Noisy & 1018.62 & 983 & 48 \\ \midrule
        \multirow{2}{*}{D5} & Quiet & 1611.70 & 1911 & 69 \\
                           & Noisy & 903.26 & 1250 & 48 \\ \midrule
        \multirow{2}{*}{D6} & Quiet & 952.64 & 1167 & 76 \\
                           & Noisy & 1028.56 & 1201 & 89 \\ \midrule
        \multirow{2}{*}{D7} & Quiet & 742.69 & 777 & 34 \\
                           & Noisy & 583.13 & 757 & 35 \\ \midrule
        \multirow{2}{*}{D8} & Quiet & 1174.82 & 1305 & 97 \\
                           & Noisy & 649.35 & 807 & 58 \\ \midrule
        \multirow{2}{*}{D9} & Quiet & 1688.16 & 2399 & 70 \\
                           & Noisy & 1315.94 & 2037 & 84 \\ \midrule
        \multirow{2}{*}{D10} & Quiet & 1696.93 & 2257 & 76 \\
                           & Noisy & 1618.92 & 2266 & 89 \\ \midrule
        \multirow{2}{*}{D11} & Quiet & 1136.75 & 1087 & 58 \\
                            & Noisy & 519.95 & 683 & 39 \\ \midrule
        \multirow{2}{*}{D12} & Quiet & 1304.85 & 1371 & 63 \\
                            & Noisy & 1185.19 & 998 & 52 \\ \midrule
        \multirow{2}{*}{D13} & Quiet & 971.70 & 874 & 63 \\
                            & Noisy & 1167.12 & 942 & 76 \\ \midrule
        \multirow{2}{*}{D14} & Quiet & 798.33 & 1034 & 43 \\
                            & Noisy & 316.04 & 484 & 17 \\ \midrule
        \multirow{2}{*}{D15} & Quiet & 817.94 & 1092 & 62 \\
                            & Noisy & 773.53 & 949 & 58 \\ \midrule
        \multirow{2}{*}{D16} & Quiet & 442.34 & 325 & 18 \\
                            & Noisy & 370.64 & 308 & 9 \\ \midrule
        \multirow{2}{*}{D17} & Quiet & 1022.70 & 1132 & 47 \\
                            & Noisy & 1198.47 & 1236 & 50 \\ \midrule
        \multirow{2}{*}{D18} & Quiet & 1105.16 & 1657 & 58 \\
                            & Noisy & 1750.69 & 2364 & 91 \\ 
        \bottomrule
    \end{tabular}
\end{table}

\newpage

\section{\add{NON-INGESTIVE ACTIVITY STRESS TEST}}

\label{app:non_ingestive_stress_test}

\add{To provide task-level detail for the practical reliability discussion in Section~\ref{sec:confound_stress_test}, we conducted an additional non-ingestive negative-control stress test. We recruited 10 participants (6 female, 4 male; ages 21--31, $M=24.60$, $SD=3.34$). Participants wore the sensing earbud and completed continuous no-eating blocks covering walking-related activities, speech and social activities, head motion and phone checking, and audio playback. Participants were instructed not to eat, chew gum, or intentionally chew during these sessions. Therefore, the entire recording served as non-chewing ground truth, and every detected chewing event was counted as a false positive.}

\add{We applied the trained chewing-detection pipeline without retraining or adding these activities as negative examples. For each task, we report false chewing detections per minute (FP/min), computed as the number of detected chewing events divided by the valid recording duration. We also report the median and interquartile range (IQR) of session-level FP/min values, together with the number of sessions with zero false positives. Category-level statistics aggregate session-level FP/min values across tasks within the same activity category.}

\begin{table}[h]
\centering
\caption{\add{False chewing detections under non-ingestive confound stress tests. All detected chewing events were counted as false positives. Category-level statistics aggregate the session-level FP/min values within each category.}}
\scriptsize
\setlength{\tabcolsep}{3.5pt}
\resizebox{\textwidth}{!}{%
\begin{tabular}{lcccccccc}
\toprule
Category 
& Total min 
& FP/min, median [IQR] 
& Category zero-FP 
& Task 
& Sessions 
& \makecell{False chewing \\ detections per min}
& FP/min, median [IQR] 
& Task zero-FP \\
\midrule

\multirow{3}{*}{Walking}
& \multirow{3}{*}{56.12}
& \multirow{3}{*}{0.63 [0.00, 1.93]}
& \multirow{3}{*}{10/30}
& Normal walking
& 10
& 0.196
& 0.00 [0.00, 0.36]
& 7/10 \\

& 
& 
&
& Light jogging
& 10
& 3.222
& 0.73 [0.12, 3.49]
& 3/10 \\

& 
& 
&
& Stair walking
& 10
&3.686
& 1.62 [0.70, 4.23]
& 0/10 \\

\midrule

\multirow{3}{*}{Speech/social}
& \multirow{3}{*}{117.38}
& \multirow{3}{*}{0.35 [0.00, 2.91]}
& \multirow{3}{*}{12/30}
& Loud reading
& 10
& 3.851
& 2.91 [0.45, 6.40]
& 2/10 \\

& 
& 
&
& Listening to another speaker
& 10
& 0.000
& 0.00 [0.00, 0.00]
& 10/10 \\

& 
& 
&
& Dyadic conversation
& 10
& 2.474
& 1.34 [0.34, 3.47]
& 0/10 \\

\midrule

\multirow{2}{*}{Head motion}
& \multirow{2}{*}{21.35}
& \multirow{2}{*}{0.00 [0.00, 0.00]}
& \multirow{2}{*}{16/20}
& Continuous head motion
& 10
& 1.544
& 0.00 [0.00, 1.73]
& 6/10 \\

& 
& 
&
& Phone checking/reading
& 10
& 0.000
& 0.00 [0.00, 0.00]
& 10/10 \\

\midrule

Playback
& 15.72
& 0.00 [0.00, 0.00]
& 10/10
& Audio playback
& 10
& 0.000
& 0.00 [0.00, 0.00]
& 10/10 \\

\bottomrule
\end{tabular}%
}
\label{tab:non_ingestive}
\end{table}

\section{GENERATION PROMPT (GPT-4O)}

\label{appendix:inter_content}

This appendix documents the generation prompt used with GPT-4o.

\begin{quote}
I currently have three theory-based intervention content designs: (1) "A bit fast" based on Dual System Theory (System 1); (2) "You are xxx chews away from reaching the recommended 25 chews" based on Control Theory; and (3) "Eating more slowly and chewing more can improve digestion and increase satiety, helping you reach your weight-loss goal sooner" based on Goal Framing Theory using a gain frame. Please help me generate sentences for these three types, and for each type include three sentence lengths: a short sentence (within 10 Chinese characters), a medium sentence (10-20 Chinese characters), and a long sentence (more than 20 Chinese characters). Thank you.
\end{quote}

\newpage

\section{SELF-REPORTED EATING EXPERIENCE QUESTIONNAIRE AND FACTOR MAPPING}

\label{appendix:daily_questionnaire}

All the questionnaire questions were asked using a 7-point Likert scale (1 = strongly disagree, 7 = strongly agree). \textit{Behavioral Positive}, \textit{Appetite}, \textit{Fullness}, and \textit{Comfort} represent four theoretically derived factors. Q6 was reverse-coded before reliability and significance analysis.

\begin{table}[h]
\centering
\caption{Mapping of questionnaire items (Q1-Q6) to factors in the Self-Reported Eating Experience scale.}
\begin{tabular}{p{0.8cm} p{9.5cm} p{3.5cm}}
\toprule
\textbf{ID} & \textbf{Item} & \textbf{Factor} \\
\hline
\hline
Q1 & I felt that my eating pace became slower. & Behavioral Positive \\
Q2 & I felt that I chewed more times per bite. & Behavioral Positive \\
Q3 & I felt that my portion size became smaller. & Appetite \\
Q4 & I felt that I became full more easily. & Fullness \\
Q5 & I felt that the intervention had a positive impact on my eating process. & Behavioral Positive \\
Q6 & When wearing earbuds during eating, I felt more uncomfortable than usual. & Comfort \\
\bottomrule
\end{tabular}

\end{table}

\section{SEMI-STRUCTURED INTERVIEW GUIDE}

\label{appendix:interview}

Interviews were semi-structured. The interviewer used the questions below as a guide and asked follow-up questions as needed. (Interviews were conducted in Chinese and translated to English for reporting.)

\subsection{Baseline Interview (Day 1)}
\begin{enumerate}
    \item \textbf{Typical eating routine.} Could you walk me through a typical lunch/dinner in your daily life (where, with whom, and how long it usually takes)?

    \item \textbf{Perceived fast-eating episodes.} In the past week, how often did you feel you ate too fast? What situations triggered it?

    \item \textbf{Drivers of fast eating.} What do you think are the main reasons you tend to eat fast?

    \item \textbf{Consequences.} When you eat fast, what changes do you notice (e.g., eating more than intended, satiety, discomfort)?

    \item \textbf{Prior attempts to slow down.} Have you tried to intentionally slow down your eating pace before? If so, what strategies did you use, and did they work?

    \item \textbf{Acceptable interventions.} If a system provides reminders/feedback during meals, what forms of prompting would you find acceptable?

    \item \textbf{When not to interrupt.} In what situations would you \emph{not} want to receive any intervention during eating?

\end{enumerate}

\subsection{Post-study Interview (Day 13)}
\begin{enumerate}
    \item \textbf{Contextual confounds.} During the study, were there any unusual events or constraints that affected your eating behavior?

    \item \textbf{Overall perceived change.} Looking back across the study, do you feel your eating pace changed?

    \item \textbf{Comparison across conditions.} Please compare your experiences across the \emph{baseline}, \emph{control}, and \emph{experimental (closed-loop)} phases.

    \item \textbf{Retention days.} On the retention days (Days 7 and 13) when the system delivered no audio prompts or feedback, how did your eating behavior compare to the days immediately before?

    \item \textbf{Control condition experience.} In the control condition , did you notice any changes in your eating pace, attention, or intake?

    \item \textbf{Closed-loop intervention experience.} In the experimental condition,  with in-meal JIT interventions and post-meal summary report, did you notice any changes in your eating pace, attention, or intake?

    \item \textbf{Perceived effects of JIT prompts.} Which in-meal JIT audio prompts (or types of prompts) had the most noticeable positive effect on you?
    
    \item \textbf{Negative effects and burden.} Did you experience any negative reactions or unintended effects from the interventions?
    
    \item \textbf{Design suggestions.} Do you have any suggestion for the system and intervention design?

\end{enumerate}

\section{PARTICIPANT-LEVEL EATING PACE TRAJECTORIES}
\label{appendix:eating_pace}


This appendix visualizes participant‑level eating‑pace trajectories derived from Earinter on-device inference logs to make individual variability explicit.
For consistency and to avoid selection bias, we fixed the same reference meal in the second day of the experiment condition. Also, we included all participants with valid data, regardless of meal choice or other factors reported in semi-interview sessions. The x‑axis indexes time‑ordered segments within a single meal, and dashed lines indicate the within‑meal median chews‑per‑swallow (CPS). As expected in in‑the‑wild settings, trajectories vary substantially: some participants show more stable or higher CPS over the meal, while others appear noisier, often when foods required minimal chewing or due to habitual chewing styles. These panels are descriptive mainly for the logic and performance of Earinter using in wild contexts.

\begin{figure*}[t]
    \centering
    \begin{subfigure}{0.49\textwidth}
        \centering
        \includegraphics[width=\linewidth]{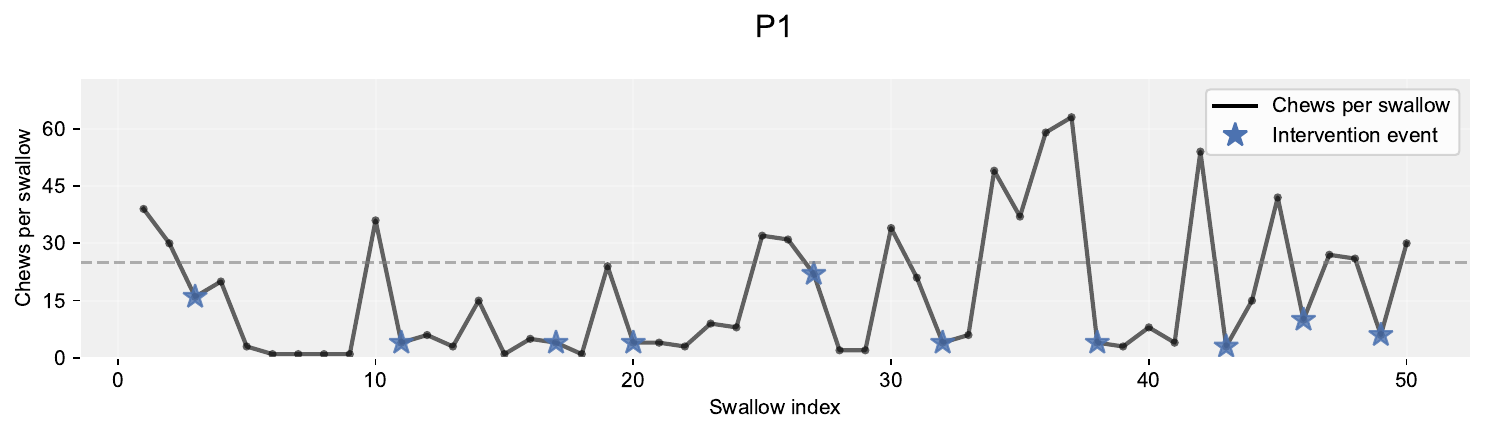}
        \label{fig:pace_P1}
    \end{subfigure}
    \hfill
    \begin{subfigure}{0.49\textwidth}
        \centering
        \includegraphics[width=\linewidth]{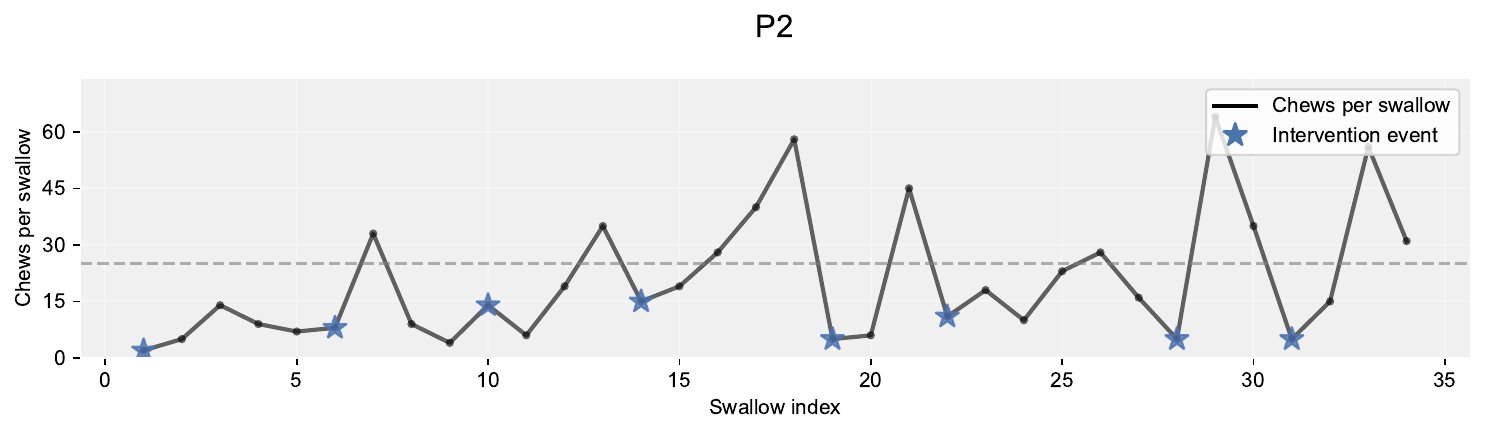}
        \label{fig:pace_P2}
    \end{subfigure}

    \begin{subfigure}{0.49\textwidth}
        \centering
        \includegraphics[width=\linewidth]{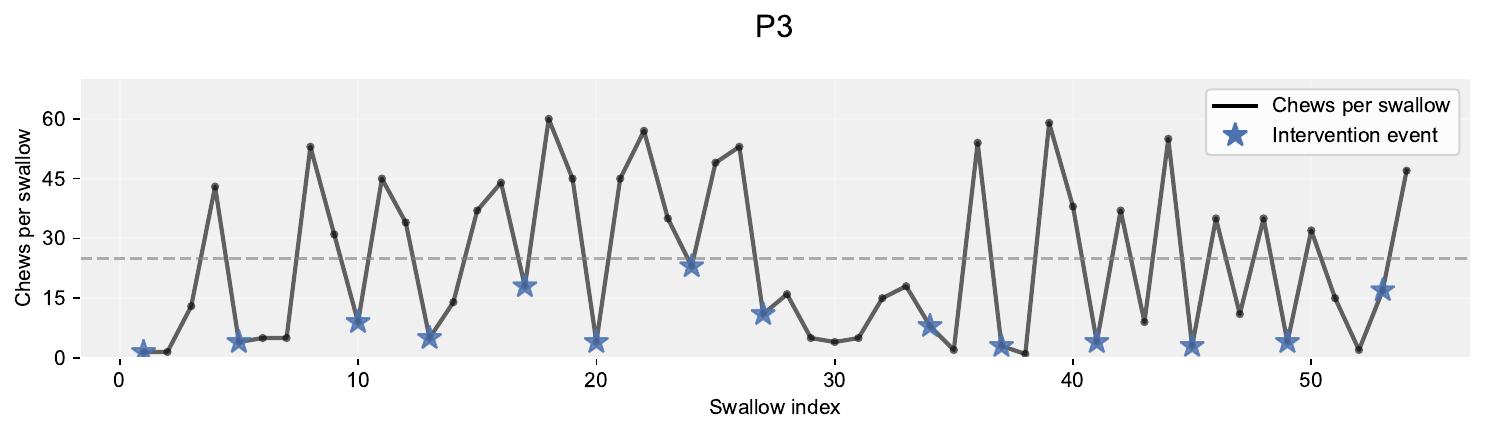}
        \label{fig:pace_P3}
    \end{subfigure}
    \hfill
    \begin{subfigure}{0.49\textwidth}
        \centering
        \includegraphics[width=\linewidth]{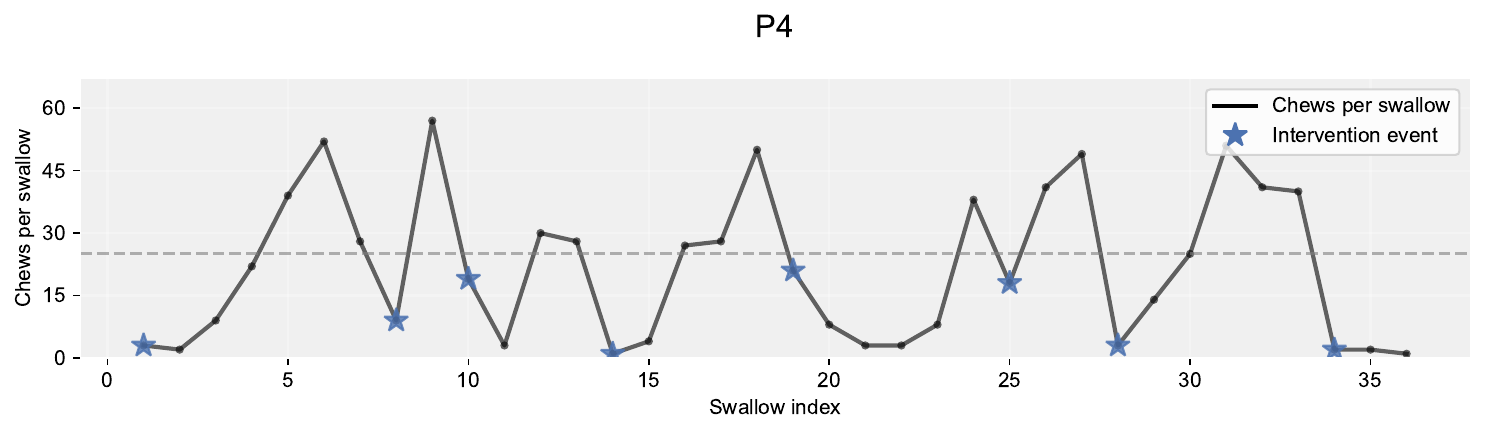}
        \label{fig:pace_P4}
    \end{subfigure}

    \begin{subfigure}{0.49\textwidth}
        \centering
        \includegraphics[width=\linewidth]{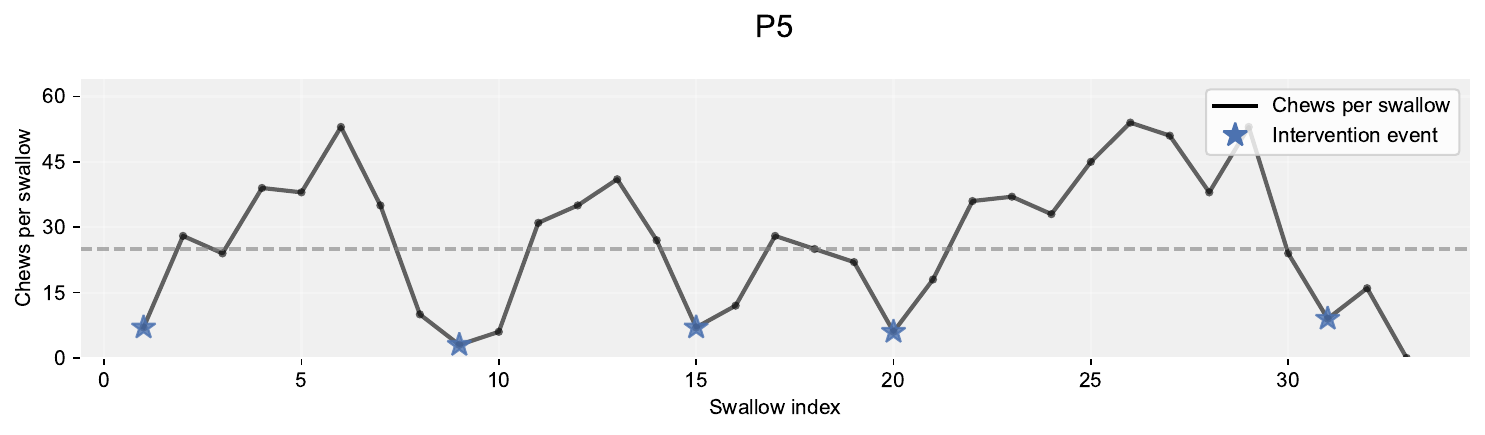}
        \label{fig:pace_P5}
    \end{subfigure}
    \hfill
    \begin{subfigure}{0.49\textwidth}
        \centering
        \includegraphics[width=\linewidth]{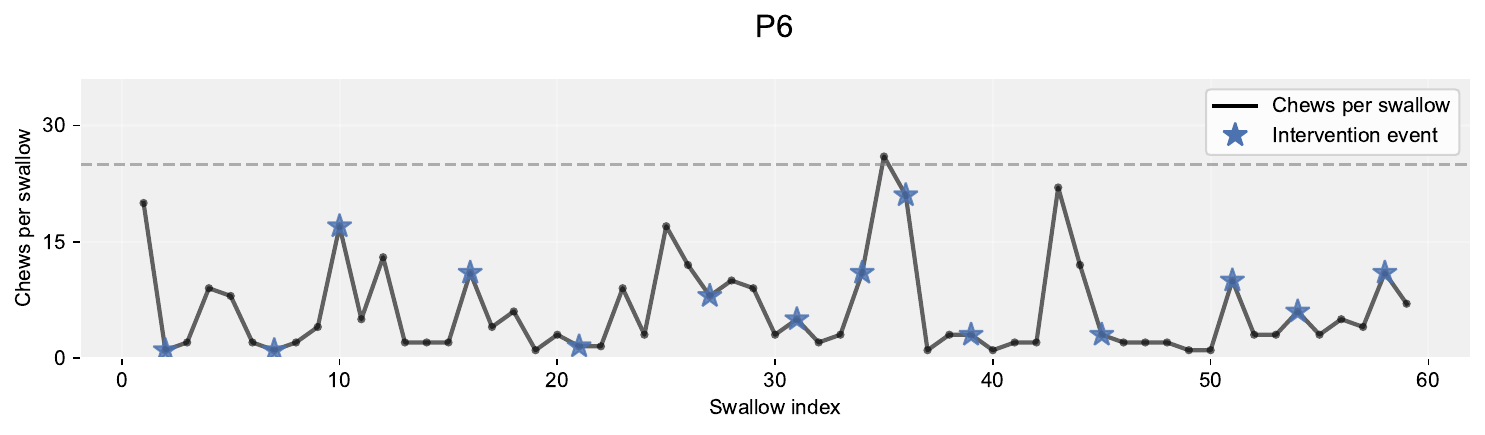}
        \label{fig:pace_P6}
    \end{subfigure}
    
    \begin{subfigure}{0.49\textwidth}
        \centering
        \includegraphics[width=\linewidth]{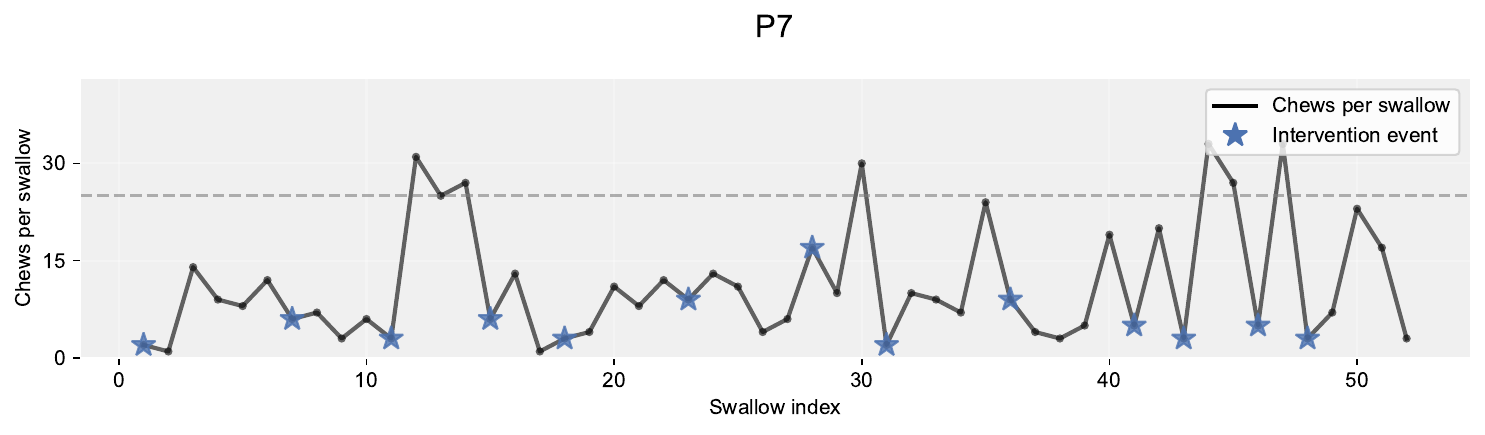}
        \label{fig:pace_P7}
    \end{subfigure}
    \hfill
    \begin{subfigure}{0.49\textwidth}
        \centering
        \includegraphics[width=\linewidth]{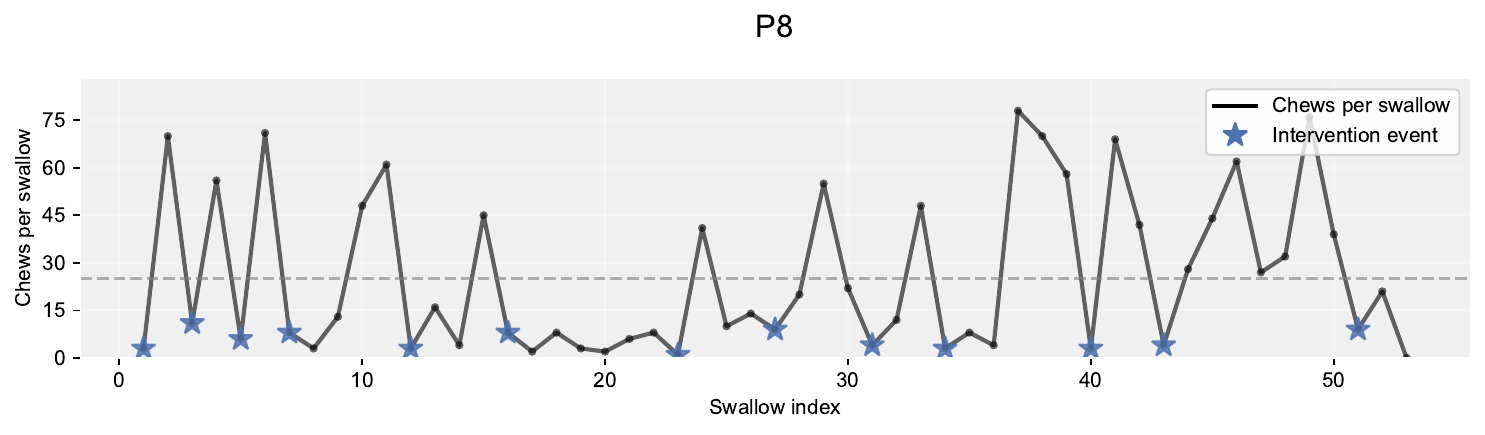}
        \label{fig:pace_P8}
    \end{subfigure}
    
    \begin{subfigure}{0.49\textwidth}
        \centering
        \includegraphics[width=\linewidth]{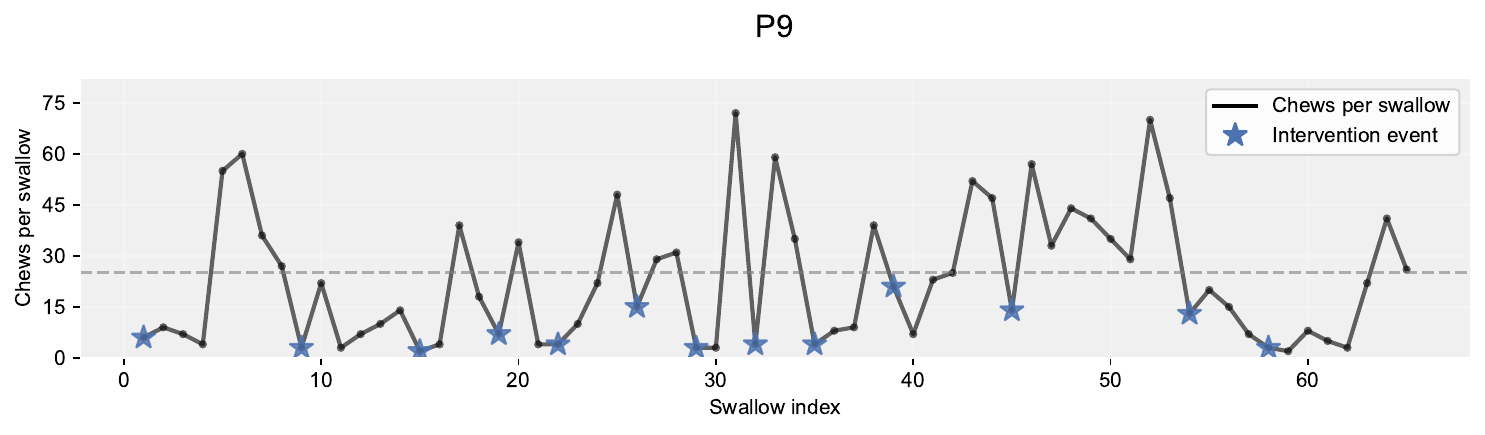}
        \label{fig:pace_P9}
    \end{subfigure}
    \hfill
    \begin{subfigure}{0.49\textwidth}
        \centering
        \includegraphics[width=\linewidth]{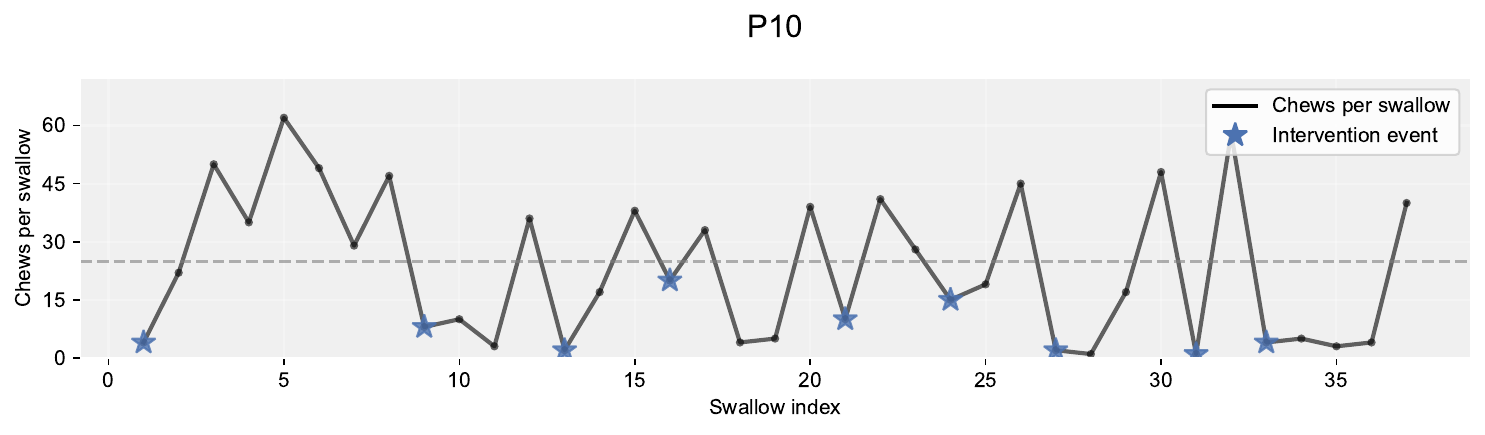}
        \label{fig:pace_P10}
    \end{subfigure}
    
    \begin{subfigure}{0.49\textwidth}
        \centering
        \includegraphics[width=\linewidth]{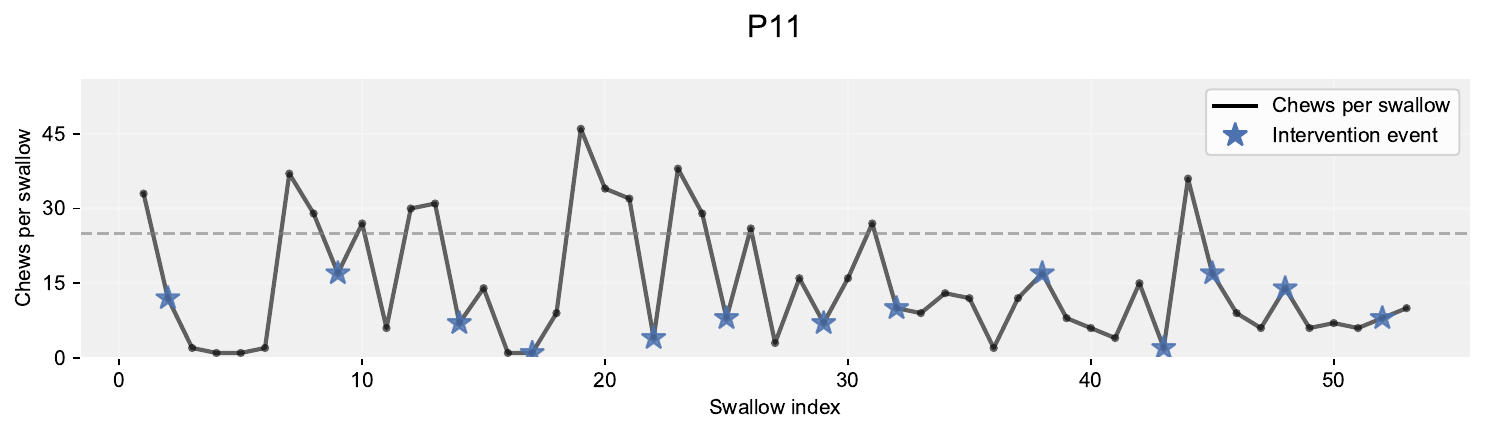}
        \label{fig:pace_P11}
    \end{subfigure}
    \hfill
    \begin{subfigure}{0.49\textwidth}
        \centering
        \includegraphics[width=\linewidth]{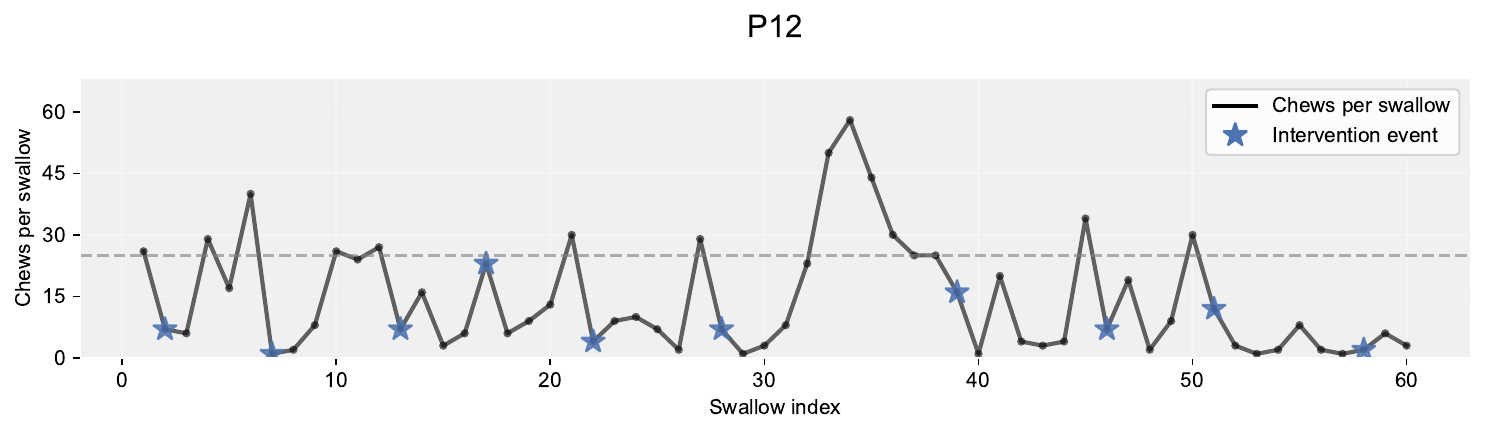}
        \label{fig:pace_P12}
    \end{subfigure}
    
    \begin{subfigure}{0.49\textwidth}
        \centering
        \includegraphics[width=\linewidth]{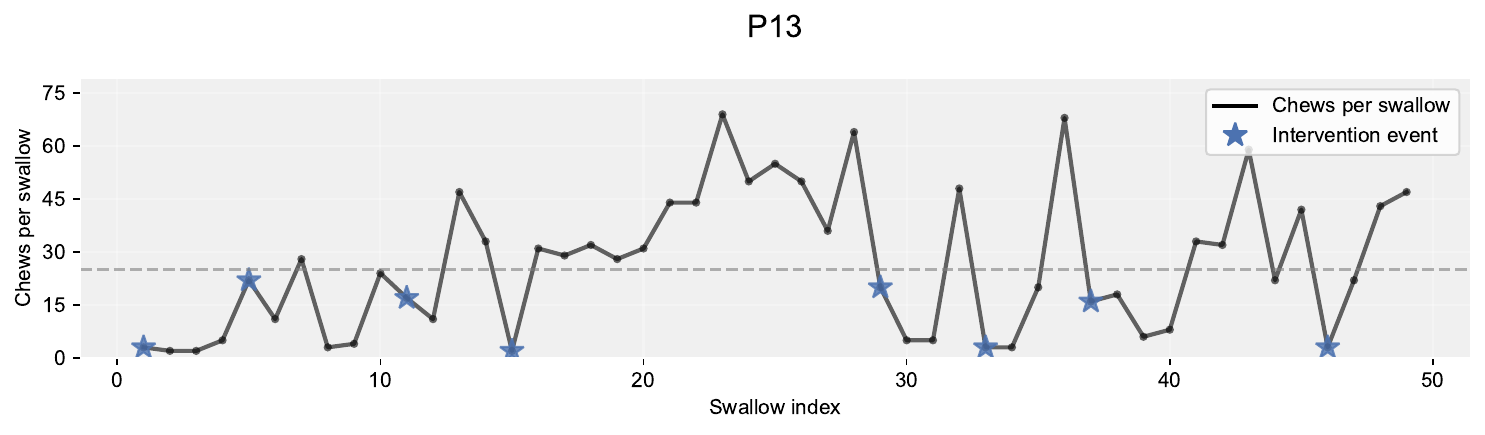}
        \label{fig:pace_P13}
    \end{subfigure}
    \hfill
    \begin{subfigure}{0.49\textwidth}
        \centering
        \includegraphics[width=\linewidth]{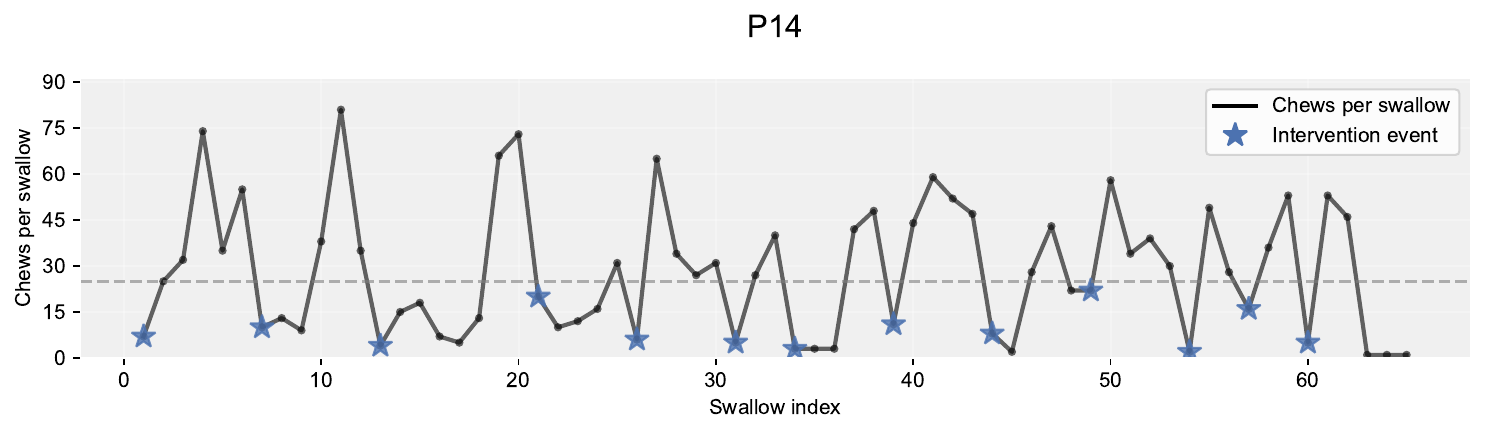}
        \label{fig:pace_P14}
    \end{subfigure}
    \caption{Participant-level trajectories of eating pace (chews-per-swallow) during a representative Experiment meal.}
    \label{fig:all_participants_same_day}
\end{figure*}

\end{document}